\documentclass[journal]{IEEEtran}

\usepackage{ifthen}
\usepackage[cmex10]{amsmath}
\usepackage{color}
\usepackage{subfigure}
\usepackage{amsmath}
\usepackage{mathtools}
\usepackage{amssymb}
\usepackage{marvosym}
\usepackage{type1cm}
\usepackage{epic}
\usepackage{wasysym}
\usepackage{latexsym}
\usepackage{graphicx}
\usepackage{placeins}
\usepackage{subfigure}
\usepackage{url}
\usepackage{fancyhdr}
\usepackage{wrapfig}
\usepackage{cite}
\usepackage{url}
\usepackage{psfrag}
\usepackage{enumerate}
\usepackage{textcomp}
\newcommand{\set}[1]{\mathcal{#1}}   
\newcommand{\cset}[1]{\mathcal{#1}^{\textnormal{c}}} 
\newcommand{\wt}[1]{\widetilde{#1}}   

\newtheorem{theorem}{Theorem}
\newtheorem{example}{Example}
\newtheorem{remark}{Remark}
\newtheorem{lemma}{Lemma}

\begin{document}
 
\title{Short Message Noisy Network Coding\\with a Decode-Forward Option}

\IEEEoverridecommandlockouts

 \author{
\IEEEauthorblockN{Jie Hou and Gerhard Kramer}\\
\IEEEauthorblockA{Institute for Communications Engineering\\
   Technische Universit\"at M\"unchen, 80290 Munich, Germany\\
       Email: \{jie.hou, gerhard.kramer\}@tum.de}
{\thanks{Version: August 09, 2013. This paper was presented in part at the International Workshop on Multi-Carrier Systems and 
Solutions, Herrsching, Germany, May 2011, at the IEEE Information Theory Workshop, Paraty, Brazil, Oct. 2011 and at the IEEE 
International Symposium on Information Theory, Boston, USA, July 2012. J. Hou and G. Kramer were supported by an
Alexander von Humboldt Professorship endowed by the German Federal Ministry of Education and Research. G. Kramer was also
supported by NSF Grant CCF-09-05235.}
}
}

\maketitle

\markboth{Submitted to IEEE Transactions on Information Theory, August 2013}{Hou and Kramer}

\begin{abstract}
Short message noisy network coding (SNNC) differs from long message noisy network coding (LNNC) in that one transmits many short 
messages in blocks rather than using one long message with repetitive encoding. Several properties of SNNC are developed. First,
SNNC with backward decoding achieves the same rates as SNNC with offset encoding and sliding window decoding for memoryless 
networks where each node transmits a multicast message. The rates are the same as LNNC with joint decoding. Second, SNNC 
enables early decoding if the channel quality happens to be good. This leads to mixed strategies that unify the advantages of 
decode-forward and noisy network coding. Third, the best decoders sometimes treat other nodes' signals as noise and an 
iterative method is given to find the set of nodes that a given node should treat as noise sources.


\end{abstract}

\begin{keywords}
Capacity, network coding, relaying.

\end{keywords}

\section{Introduction}
{\em Noisy Network Coding} (NNC) extends network coding from noiseless to noisy networks. NNC is based on the compress-forward 
(CF) strategy of \cite{Cover01} and there are now two {\em encoding} variants: short message NNC (SNNC) \cite{Schein01, 
Kramer04, Ya01, Ya02, Ya03, Xie02, GH01, GH02, Buddy01, Hou03, Pramod01} and long message NNC (LNNC)\cite{Suhas01, Lim01, 
Gamal01}. Both variants achieve the same rates that include the results of \cite{Yeung01, Effros01,  Kramer05} as special cases.

For SNNC, there are many {\em decoding} variants: step-by-step decoding \cite{Cover01, Schein01, Kramer04, Ya01}, 
sliding window decoding \cite{Ya02,Ya03}, backward decoding \cite{Xie02,GH01, 
GH02, Buddy01, Hou03} and joint decoding \cite{Buddy01}. There are also several {\em initialization} methods. The 
papers \cite{Ya01, Ya02, Ya03} use delayed (or offset) encoding, 
\cite{Xie02} uses many extra blocks to decode the last quantization messages and \cite{Hou03} uses extra blocks to transmit the 
last quantization messages by multihopping. We remark that the name of the relaying operation should not depend on which {\em 
decoder} (step-by-step, sliding window, joint, or backward decoding) is used at the {\em destination} but is a generic name for 
the {\em processing} at the {\em relays}, or in the case of SNNC and LNNC, the overall encoding strategy of the network nodes. 


More explicitly, SNNC has
\begin{itemize}
\item \textbf{Sources} transmit independent short messages in blocks.
\item \textbf{Relays} perform CF but perhaps without hashing (or binning) which is called quantize-forward (QF).
\item \textbf{Destinations} use one of the several decoders. For instance, SNNC with CF and step-by-step decoding was studied for 
relay networks in \cite[Sec. 3.3.3]{Schein01}, \cite[Sec. V]{Kramer04}, and \cite{Ya01}. The papers \cite{Ya02, Ya03} studied 
SNNC with {\em sliding window} decoding. The papers \cite{Xie02, GH01, GH02, Buddy01, Hou03} considered SNNC with {\em backward} 
decoding. SNNC with joint decoding was studied in \cite{Buddy01}.
\end{itemize}
We prefer backward decoding because it permits {\em per-block} processing and gives the most direct way of establishing rate 
bounds. However, we remark that the sliding window decoder of \cite{Ya02, Ya03} is preferable because of its lower decoding 
delay, and because it enables streaming.

LNNC uses three techniques from \cite{Suhas01}:
\begin{itemize}
\item \textbf{Sources} use repetitive encoding with {\em long} messages.
\item \textbf{Relays} use QF.
\item \textbf{Destinations} decode all messages and all quantization bits jointly.
\end{itemize}

One important drawback of long messages is that they inhibit decode-forward (DF) even if the channel
conditions are good \cite{GH01}. For example, if one relay is close to the source and has a strong source-relay link, then the
natural operation is DF which removes the noise at the relay. But this is generally not possible with a long message because of
its high rate.

The main goals of this work are to simplify and extend the single source results of \cite{Xie02, GH01, GH02} by developing SNNC 
with backward decoding for networks with {\em multiple multicast} sessions \cite{Hou03}. We also introduce the 
following methods:
\begin{itemize}
 \item Multihopping to initialize backward decoding. This method reduces overhead as compared to the joint decoder initialization 
used in \cite{Xie02}. The method further enables {\em per-block} processing for all signals, i.e., all messages and quantization 
indices.

\item An iterative proof technique to find the set of nodes that a destination should treat as noise (the same argument was 
used in \cite[Sec. IV-C]{Ya02}). 
\end{itemize}

%

This paper is organized as follows. In Section \ref{notations}, we state the problem. In Section \ref{SSNC}, we show that SNNC
achieves the same rates as SNNC with sliding window decoding and LNNC for memoryless networks with multiple multicast sessions. 
In Section \ref{diss}, we discuss the results and relate them to other work. In Section \ref{mixed}, we present coding schemes 
for mixed strategies that allow relay nodes to switch between DF and QF depending on the channel conditions. Results on Gaussian 
networks are discussed in Section \ref{Gauss}. Finally, Section \ref{ConRe} concludes the paper.

\section{Preliminaries}
\label{notations}

\subsection{Random Variables}
Random variables are written with upper case letters and their realizations with the corresponding lower
case letters. Bold letters refer to random vectors and their realizations. A random variable $X$ has distribution $P_X$. We write
probabilities with subscripts $P_X(x)$ but we drop the subscripts if the arguments of the distributions are lower case versions of
the random variables. For example, we write $P(x)=P_X(x)$. Calligraphic letters denote sets, e.g., we write $\set K=\{1,2,\dots,
K\}$. The size of a set $\set S$ is denoted as $|\set S|$ and the complement set of $\set S$ is denoted as $\cset S$. Subscripts
on a symbol denote the symbol's source and the position of the symbol in a sequence. For instance, $X_{ki}$ denotes the $i$-th
output of the $k$-th encoder. Superscripts denote finite-length sequences of symbols, e.g., $x^n_k=(x_{k1},\dots, x_{kn})$. Set
subscripts denote vectors of letters, e.g., $X_\mathcal S=[X_k: k\in \mathcal S]$. We use $\set T^n_\epsilon(P_X)$ to denote the 
set of letter-typical sequences of length $n$ with respect to the probability distribution $P_X$ and the non-negative number 
$\epsilon$ \cite[Ch. 3]{ Massey01}, \cite{Roche01 }, i.e., we have
\begin{align*}
\set T^n_\epsilon(P_X)=\left\{x^n:\Big| \frac{N(a|x^n)}{n} -P_X(a) \Big| \le \epsilon P_X(a),\; \forall a\in \set X  \right\}
\end{align*}
where $N(a|x^n)$ is the number of occurrences of $a$ in $x^n$.

\begin{figure}[t!]
\centering
\psfrag{w1}[][][0.9]{$W_1\rightarrow (X_1,Y_1)$}
\psfrag{w2}[][][0.9]{$W_2\rightarrow (X_2,Y_2)$}
\psfrag{w3}[][][0.9]{$W_3\rightarrow (X_3,Y_3)$}
\psfrag{wk}[][][0.9]{$W_k\rightarrow (X_k,Y_k)$}
\psfrag{wN}[][][0.9]{$W_K\rightarrow (X_K,Y_K)$}
\psfrag{dot1}{$\vdots$}     
\psfrag{dot2}{$\vdots$}
\psfrag{dot3}{$\cdots$}
\psfrag{p}[][][0.9]{$P(y_1,\dots,y_K|x_1,\dots,x_K)$}
\includegraphics[width=7.1cm]{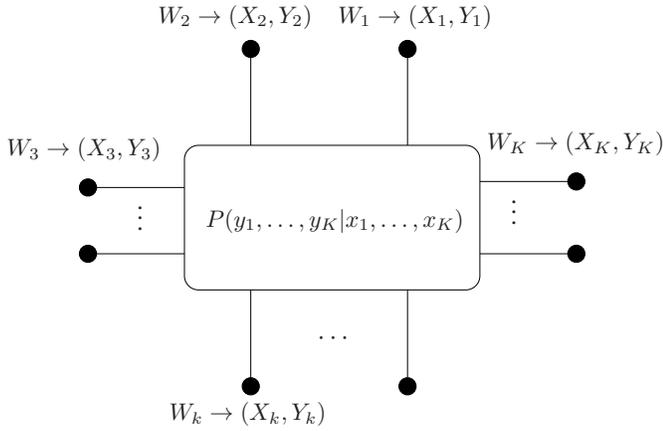}
\caption{A $K$-node memoryless network. The network is a DMN if the alphabets of $X_k$ and $Y_k$
are discrete and finite for $k=1,\dots, K$.}
\label{DMN}
\end{figure}

\subsection{Memoryless Networks}
Consider the $K$-node memoryless network depicted in Fig.~\ref{DMN} where each node has one message only. This model does not
include broadcasting messages and was used in \cite{Gamal01} and \cite[Ch. 15]{Cover03} . Node $k$, $k\in
\mathcal K$, has a message $W_k$ destined for nodes in the set $\mathcal D_k$, $\mathcal D_k \subseteq \mathcal K\setminus \{k\}$,
while acting as a relay for messages of the other nodes. We write the set of nodes whose signals node $k$ must decode correctly as
$\set {\widetilde D}_k=\{ i\in \set K:\;k\in \set D_i\}$. The messages are mutually statistically independent and $W_k$ is
uniformly distributed over the set $\{1,\dots, 2^{nR_k}\}$, where $2^{nR_k}$ is taken to be a non-negative integer.

The channel is described by the conditional probabilities
\begin{align}
\label{ChannelDistribution}
P(y^K|x^K)=P(y_1,\dots,y_K|x_1,\dots,x_K)
\end{align}
where $\mathcal X_k$ and $\mathcal Y_k$, $k\in \mathcal K$, are the respective input and output alphabets,
i.e., we have
\begin{align*}
(x_1,\dots,x_K)&\in \mathcal X_1 \times \cdots \times \mathcal X_K\\
(y_1,\dots,y_K)&\in \mathcal Y_1\times \cdots \times \mathcal Y_K.
\end{align*}
If all alphabets are discrete and finite sets, then the network is called a {\em discrete} memoryless network (DMN)
\cite{Kramer03},\cite[Ch.18]{Gamal03}. As usual, we develop our
random coding for DMNs and later extend
the results to Gaussian channels. Node $k$ transmits $x_{ki} \in \mathcal X_k$ at time $i$ and receives $y_{ki} \in \mathcal Y_k$.
The channel is {\em memoryless} and {\em time invariant} in the sense that
\begin{align}
P(&y_{1i},\dots,y_{Ki} | w_1, \dots, w_K, x^i_1,\dots, x^i_{K}, y^{i-1}_1,\dots,y^{i-1}_{K})\notag\\
&=P_{Y^K|X^K}(y_{1i},\dots, y_{Ki}|x_{1i},\dots, x_{Ki})
\end{align}
for all $i$. 

\begin{table*}[t!]
\begin{center}
\begin{tabular}{r|c c c  c}
  \hline
  Block   & 1                                      &       $\cdots$   &            $B$ &           $B+1$  $\cdots$ 
$B+K\cdot(K-1)$\\
\hline
$X_1$ & $\mathbf x_{11}(w_{11} , 1)$      &  $\cdots$   &  $\mathbf
x_{1B}(w_{1B},l_{1(B-1)})$   &  \\

$\hat Y_1$  & $\mathbf {\hat y}_{11}(l_{11}|w_{11}, 1)$ &  $\cdots$   & 
$\mathbf {\hat y}_{1B}(l_{1B}|w_{1B}, l_{1(B-1)})$  &  \\

$\vdots$& $\vdots$& $\vdots$&$\vdots$&Multihop $K$ messages to $K-1$ nodes\\

$X_K$  & $\mathbf x_{K1}(w_{K1} , 1)$   & $\cdots$    &  $\mathbf
x_{KB}(w_{KB},l_{K(B-1)})$& in $K\cdot (K-1)\cdot n^\prime$ channel uses\\

$\hat Y_K$  & $\mathbf {\hat y}_{K1}(l_{K1}|w_{K1}, 1)$ &   & 
$\mathbf {\hat y}_{KB}(l_{KB}|w_{KB}, l_{K(B-1)})$  & \\
\hline
\end{tabular}
\end{center}
\caption{SNNC for one multicast session per node.}
\label{encoding}
\end{table*}

\subsection{Flooding}
We can represent the DMN as a directed graph $\mathcal G=\{\mathcal K, \mathcal E \}$, where $\mathcal E
\subset \mathcal K \times \mathcal K$ is a set of edges. Edges are denoted as $(i,j) \in \mathcal E,\; i,j \in
\mathcal K, \;i\ne j$. We label edge $(i,j)$ with the non-negative real number
\begin{align}
C_{ij}=\underset{x_{\set K\setminus i}}{\text{max}}\; \underset{P_{X_i}}{\text{max}}\;I(X_i;Y_j|X_{\mathcal K\setminus
i}=x_{\mathcal K\setminus i} )
\end{align}
called the capacity of the link, where $I(A;B|C=c)$ is the mutual information between the random variables $A$ and $B$
conditioned on the event $C=c$. Let $\text{Path}_{(i,j)}$ be a path that starts from node $i$ and ends at node $j$. Let
$\Gamma_{(i,j)}$ to be the set of such paths. We write $(k,\ell) \in \text{Path}_{(i,j)}$ if $(k,\ell)$
lies on the path $\text{Path}_{(i,j)}$. We may communicate reliably between
nodes $i$ and $j$ if 
\begin{align}
R_{ij}=\underset{ \text{Path}_{(i,j)} \in \Gamma_{(i,j)} }{\text{max}}
\;\underset{(k,l)\in \text{Path}_{(i,j)}}{\text{min}}\;C_{kl}
\end{align}
is positive. We assume that $R_{ij}>0$ for all nodes $i$ with a message destined for node $j$. Observe that if $C_{ij}>0$ for
all $i,j$, then at most $K-1$ hops are needed for node $i$ to reliably convey its message at rate
\begin{align}
\underset{j\in \set K}{\text{min}}\; R_{ij}
\end{align}
by multihopping to all other nodes in the network. Hence, for a $K$-node memoryless network at most $K(K-1)
\label{multihopping}$ hops are needed for all nodes to \lq\lq flood\rq\rq\;their messages by multihopping through the network.

\begin{example}
\label{exp:example1}
\begin{figure}[t!]
\centering
\includegraphics[width=6.8cm]{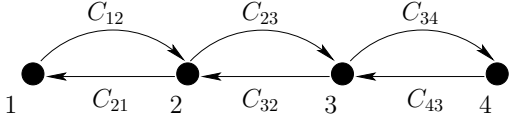}
\caption{A line network with $4$ nodes. Each node can communicate reliably with any other node as long as $C_{ij}>0$ for all
$i,j$. }
\label{lineNet}
\end{figure}

A line network with $4$ nodes is depicted in Fig.~\ref{lineNet}. Node $1$ has a message for node $4$ and we assume
that $C_{12}>0$, $C_{23}>0$ and $C_{34}>0$ so that node $1$ can communicate reliably to node $4$ by multihopping through nodes $2$
and $3$ with $3$ hops.
\end{example}

\subsection{Encoders and Decoders}
We define two types of functions for each node $k$:
\begin{itemize}
 \item $n$ encoding functions $f^n_k=(f_{k1},\dots,f_{kn})$ that generate channel inputs based on the local message and past
channel outputs
\begin{align}
 X_{ki}=f_{ki}(W_{k}, Y^{i-1}_{k}), \;i=\{1,\dots, n\}.
\end{align}
\item One decoding function 
\begin{align}
g_k(Y^n_k,W_k)=[\hat W^{(k)}_i, \;i \in \mathcal {\widetilde D}_k]
\end{align}
where $\hat W^{(k)}_i$ is the estimate of $W_i$ at node $k$.
\end{itemize}

The average error probability for the network is defined as
\begin{align}
&P^{(n)}_e=\text{Pr}\left[ \bigcup_{k\in \set K} \bigcup_{i\in {\mathcal {\widetilde D}_k}} \{\hat W^{(k)}_i \ne W_i\}\right].
\end{align}
A rate tuple $(R_1,\dots,R_K)$ is achievable for the DMN if for any $\epsilon >0$, there is a sufficiently large integer $n$ and
some functions $\{f^n_k\}^K_{k=1}$ and $\{g_k\}^{K}_{k=1}$ such that $P^{(n)}_e\le \epsilon$. The capacity region is the closure
of the set of achievable rate tuples. For each node $k$ we define
\begin{align}
\set K_k&=\{k\} \cup \set {\widetilde D}_k \cup \set T_k,\;\set T_k\subseteq \cset {\widetilde D}_k\setminus \{k\} \label{KK}
\end{align}
where $\set T_k$ has the nodes whose messages node $k$ is not interested in but whose symbol sequences are included in the
typicality test in order to remove interference. We further define, for any $\set S\subset \set L\subseteq \set K$, the quantities
\begin{align}
&I^{\set L}_{\set S}(k)= I(X_{\mathcal S}; \hat Y_{\cset S} Y_k |X_{\cset S})-I(\hat Y_{\mathcal S}; Y_{\mathcal
S}|X_{\set L} \hat Y_{\cset S} Y_k)\label{RateNoTime}\\
&I^{\set L}_{\set S}(k|T)= I(X_{\set S}; \hat Y_{\cset S} Y_k |X_{\cset S}T)-I(\hat
Y_{\mathcal S}; Y_{\mathcal S}|X_{\set L} \hat Y_{\cset S} Y_kT)\label{RateTime}
\end{align}
where $\cset S$ in (\ref{RateNoTime}) and (\ref{RateTime}) is the complement of $\set S$ in $\set L$. We write $R_{\set
S}=\sum_{k\in \set S} R_k$. 

\section{Main Result and Proof}
\label{SSNC}

The following theorem is the main result of this paper.
\begin{theorem}
\label{thm:theorem1}
For a $K$-node memoryless network with one multicast session per node, SNNC with backward decoding achieves the same 
rate tuples $(R_1,\dots,R_K)$ as SNNC with sliding window decoding \cite{Ya02, Ya03} and LNNC with joint decoding 
\cite{Gamal01, Lim01}. These are the rate tuples satisfying
\begin{align}
0\le R_{\set S}< \;I^{\set K_k}_{\set S}(k|T)\label{Theo1}
\end{align}
for all $k\in \set K$, all subsets $\set S \subset \set K_k$ with $k \in \cset S$ and $\set S \cap \wt {\mathcal D}_k \ne
\emptyset$, where $\cset S$ is the complement of $\set S$ in $\set K_k$, and for
joint distributions that factor as
\begin{align}
\label{distribution1}
P(t)&\left[\prod^K_{k=1}P(x_k| t)P(\hat y_k| y_k, x_k,t)\right]P(y^K|x^K).
\end{align}
\end{theorem}
\begin{remark}
\label{rem:remark1}
The set $\set K_k$ (see (\ref{KK})) represents the set of nodes whose messages are known or decoded at node $k$. In other words,
from node $k$'s perspective the network has nodes $\set K_k$ only.
\end{remark}
\begin{example}
\label{exp:example2}
If $\mathcal D=\mathcal D_1=\dots=\mathcal D_K$, then the bound (\ref{Theo1}) is taken for all $k\in \set K$ and all
subsets $\set S \subset \set K_k$ with $k \in \cset S$ and $\set S \cap \set D \ne \emptyset$, where $\cset S$ is the
complement of $\set S$ in $\set K_k$.
\end{example}

\begin{example}
\label{exp:example3} 
Consider $\set K=\{1,2,3,4\}$ and suppose node $1$ has a message destined for node $3$, and node $2$ has a message destined
for node $4$. We then have $\set {\wt D}_3=\{1\}$ and $\set {\wt D}_4=\{2\}$. If nodes 3 and 4 choose $\set T_3=\{2\}$ and $\set
T_4=\{\emptyset\}$ respectively, then we have $\set K_3=\{1,2,3\}$ and $\set K_4=\{2,4\}$. In this case the rate bounds
(\ref{Theo1}) are:\\
Node $3$:
\begin{IEEEeqnarray}{rCl}
R_1&<& I(X_1;\hat Y_2\hat Y_3Y_3|X_2X_3 T)\\\
R_1+R_2&<& I(X_1X_2;\hat Y_3Y_3|X_3 T)\notag\\
  &&-I(\hat Y_1\hat Y_2;Y_1Y_2|X_1X_2X_3\hat Y_3Y_3T) 
\end{IEEEeqnarray}
Node $4$:
\begin{align}
R_2&<I(X_2;\hat Y_4Y_4|X_4 T)-I(\hat Y_2;Y_2|X_2X_4\hat Y_4Y_4T)
\end{align}
\end{example}

\subsection{Encoding}
To prove Theorem~\ref{thm:theorem1}, we choose $\set K_k=\set K$ for all $k$ for simplicity. We later discuss the case where
these sets are different. For clarity, we set the time-sharing random variable $T$ to be a constant. Table~\ref{encoding} shows
the SNNC encoding process. We redefine $R_k$ to be the rate of the short messages in relation to the (redefined) block length $n$.
In other words, the message $w_k, k\in \mathcal K$, of ${nBR_k}$ bits is split into $B$ equally sized blocks, $w_{k1},\dots,
w_{kB}$, each of $nR_k$ bits. Communication takes place over $B+K\cdot (K-1)$ blocks and the true rate of $w_k$ will be
\begin{equation}
\label{TrueRate}
R_{k,\text{true}}=\frac{nBR_k}{nB+[K\cdot (K-1)\cdot n^\prime]}
\end{equation}
where $n^\prime$ is defined in (\ref{nprime}) below.

{\em Random Code:} Fix a distribution $\prod^K_{k=1}P(x_k)P(\hat y_k| y_k, x_k)$. For each block
$j=1,\dots,B$ and node $k\in \mathcal K$, generate $2^{n(R_k+\hat R_k)}$ codewords $\mathbf x_{kj}(w_{kj}, l_{k(j-1)})$,
$w_{kj}=1,\dots, 2^{nR_k}, l_{k(j-1)}=1,\dots, 2^{n\hat R_k}$, according to $\prod^n_{i=1}P_{X_k}(x_{(kj)i})$ where $l_{k0}=1$ by
convention. For each $w_{kj}$ and $l_{k(j-1)}$, generate  $2^{n\hat R_k}$ reconstructions $\mathbf {\hat y}_{kj}(l_{kj}|w_{kj},
l_{k(j-1)})$, $l_{kj}=1,\dots, 2^{n\hat R_k}$, according to $\prod^n_{i=1}P_{\hat Y_k|X_k}(\hat y_{(kj)i}|x_{(kj)i}(w_{kj},
l_{k(j-1)}))$. This defines the codebooks
\begin{IEEEeqnarray}{rCl}
 \set C_{kj}&=\{ &\mathbf x_{kj}(w_{kj}, l_{k(j-1)}), \mathbf {\hat y}_{kj}(l_{kj}|w_{kj}, l_{k(j-1)}),\notag\\
&&w_{kj}=1,\dots, 2^{nR_k}, \;l_{k(j-1)}=1,\dots,2^{n\hat R_k},\notag\\
&&l_{kj}=1,\dots,2^{n\hat R_k}\}
\end{IEEEeqnarray}
for $j=1,\dots, B$ and $k\in \mathcal K$.

The codebooks used in the last $K(K-1)$ blocks with $j>B$ are different. The blocks 
\begin{align}
\label{extraBlocks}
j=B+(k-1)\cdot (K-1)+1,\dots,B+k\cdot (K-1)
\end{align}
are dedicated to flooding $l_{kB}$ through the network, and for all nodes $\tilde k\in\set K$ we generate $2^{n^\prime \hat
R_{k}}$ independent and identically distributed (i.i.d.) codewords $\mathbf x_{\tilde kj}(l_{kB})$, $l_{kB}=1,\dots,2^{n^\prime
\hat R_{k}}$, according to $\prod^{n^\prime}_{i=1}P_{X_{\tilde k}}(x_{(\tilde kj)i})$. We choose 
\begin{equation}
\label{nprime}
n^\prime=\underset{k}{\text{max}} \;\frac{n\hat R_k}{ \underset{\tilde k\in \mathcal K}{\text{min}}\;R_{k\tilde k}} 
\end{equation}
that is independent of $k$ and $B$. The overall rate of user $k$ is thus given by (\ref{TrueRate})
which approaches $R_k$ as $B\rightarrow \infty$.

{\em Encoding:}
Each node $k$ upon receiving $\mathbf {y}_{kj}$ at the end of block $j$, $j\le B$, tries to find an index $l_{kj}$ such
that the following event occurs:
\begin{align}
E_{0(kj)}(l_{kj}): &\left(\mathbf {\hat y}_{kj}(l_{kj}|w_{kj},l_{k(j-1)}),\; \mathbf x_{kj}(w_{kj},l_{k(j-1)}),\; \mathbf
y_{kj}\right)
\notag \\&\in \set T^{n}_\epsilon\left(P_{\hat Y_kX_kY_k}\right)
\end{align}
If there is no such index $l_{kj}$, set $l_{kj}=1$. If there is more than one, choose one. Each
node $k$ transmits $\mathbf x_{kj}(w_{kj}, l_{k(j-1)})$ in block $j=1,\dots, B$.

In the $K-1$ blocks (\ref{extraBlocks}), node $k$ conveys $l_{kB}$ reliably to all other nodes by multihopping $\mathbf x_{kj}(
l_{kB})$ through the network with blocks of length $n^\prime$.

\subsection{Backward Decoding} 
\label{backward}
Let $\epsilon_1>\epsilon$. At the end of block $B+K\cdot (K-1)$ every node $k\in \mathcal K$ has
reliably recovered $\mathbf l_B=(l_{1B},\dots,l_{KB})$ via the multihopping of the last $K(K-1)$ blocks.

For block $j=B,\dots,1$, node $k$ tries to find tuples ${\mathbf {\hat w}^{(k)}_j}=({\hat
w}^{(k)}_{1j},\dots,{\hat w}^{(k)}_{Kj})$ and $\mathbf {\hat l}^{(k)}_{j-1}=(\hat l^{(k)}_{1(j-1)},\dots,\hat l^{(k)}_{K(j-1)})$
such that the following event occurs:
\begin{align}
&E_{1({kj})}({\mathbf {\hat w}^{(k)}_j},{\mathbf {\hat l}^{(k)}_{j-1}}, {\mathbf {l}_{j}} ):\notag\\
&\left(\mathbf x_{1j}(\hat w^{(k)}_{1j}, \hat l^{(k)}_{1(j-1)}), \dots, \mathbf x_{Kj}(\hat w^{(k)}_{Kj}, \hat
l^{(k)}_{K(j-1)}),\right.\notag\\
&\left.\; \mathbf
{\hat y}_{1j}(l_{1j}|\hat w^{(k)}_{1j},\hat
l^{(k)}_{1(j-1)}), \dots,\mathbf {\hat y}_{Kj}(l_{Kj}|\hat w^{(k)}_{Kj},\hat
l^{(k)}_{K(j-1)}), \mathbf y_{kj}\right)\notag \\
&\quad \in \set T^{n}_{\epsilon_1}\left(P_{X_{\mathcal K}\hat Y_\mathcal K Y_k}\right) \label{TypCheck}
\end{align}
where $\mathbf l_j=(l_{1j},\dots,l_{Kj})$ has already been reliably recovered from the previous block $j+1$.

{\em Error Probability:} Let $\mathbf 1=(1,\dots,1)$ and assume without loss of generality that $\mathbf w_j=\mathbf 1$
and $\mathbf l_{j-1}=\mathbf 1$. In each block $j$, the error events at node $k$ are:
\begin{align}
E_{(kj)0}:&\;{\cap_{l_{kj}}}\;E^\text{c}_{0(kj)}(l_{kj})\ \label{eventblock1}\\
E_{(kj)1}:&\;E^\text{c}_{1(kj)}(\mathbf 1,\mathbf 1,\mathbf 1)\label{eventblock2}\\
E_{(kj)2}:&\;{\cup_{({\mathbf w}_{j}, \mathbf l_{j-1})\ne (\mathbf 1, \mathbf 1)
}}\;E_{1(kj)}(\mathbf w_j,\mathbf l_{j-1}, \mathbf 1) \label{eventblock3}
\end{align}
The error event  $E_{kj}=\cup_{i=0}^2E_{(kj)i}$ at node $k$ in block $j$ thus satisfies
\begin{align}
\label{UnionBound}
\text{Pr}\big[E_{kj}\big] \le \sum^2_{i=0}\text{Pr}\big[E_{(kj)i}]
\end{align}
where we have used the union bound. $\text{Pr}\big[E_{(kj)0}\big]$ can be made small with large $n$, as long as (see
\cite{Roche01})
\begin{equation}
\label{rdbound}
\hat R_k > I( \hat Y_{k}; Y_{k}|X_k) + \delta_{\epsilon}(n)
\end{equation}
where $\delta_{\epsilon}(n) \rightarrow 0$ as $n\rightarrow \infty$. Similarly, $\text{Pr}\big[E_{(kj)1}\big]$ can be made small 
with large $n$.

To bound $\text{Pr}\big[E_{(kj)2} \big]$,  for each $\mathbf w_j$ and $\mathbf l_{j-1}$ we define 
\begin{align}
\set M(\mathbf w_j)&=\{i \in \mathcal K : w_{ij}\ne 1\}\\
\mathcal Q(\mathbf l_{j-1})&=\{i \in \mathcal K :l_{i(j-1)}\ne 1\}\\
\set S(\mathbf w_j, \mathbf l_{j-1})&=\mathcal M(\mathbf w_j) \cup \mathcal Q(\mathbf l_{j-1})
\end{align}
and write $\set S=\set S(\mathbf w_j, \mathbf l_{j-1}) $. The important observations are:
\begin{itemize}
 \item $(\mathbf X_{\mathcal S}, \mathbf {\hat Y}_{\mathcal S})$ is independent of $(\mathbf X_{\cset S}, \mathbf {\hat
Y}_{\cset S}, \mathbf Y_{kj} )$ in the random coding experiment;
\item The $(X_i, {\hat Y}_i),\; i \in \set S,$ are mutually independent.
\end{itemize}
For $k\in \cset S$ and $(\mathbf w_j, \mathbf l_{j-1})\ne ( \mathbf 1, \mathbf 1)$, we thus have
\begin{align}
\label{PEISJ}
&\text{Pr}\big[E_{1(kj)}(\mathbf w_j, \mathbf l_{j-1}, \mathbf l_j)\big]\le 2^{-n(I_{\set S}-\delta_{\epsilon_1}(n))}
\end{align}
where $\delta_{\epsilon_1}(n)\rightarrow 0$ as $n\rightarrow \infty$ and 
\begin{align}
\label{ISJ}
I_{\set S}&=\left[\sum_{i\in \set S}H(X_i {\hat Y}_i)\right]+H(X_{\cset S} {\hat Y}_{\cset S}
Y_k)\notag\\&\quad\quad\quad\quad-H(X_{\mathcal S}{\hat Y}_{\mathcal S} X_{\cset S} {\hat Y}_{\cset S} Y_k)\notag \\
&=I(X_{\mathcal S}; \hat Y_{\cset S} Y_k| X_{\cset S})\notag\\&\quad\quad\quad+\left[\sum_{i\in\mathcal S} H(\hat
Y_i|X_i)\right]-H(\hat Y_{\set S}|X_{\set K}\hat Y_{\cset S}Y_k).
\end{align}
By the union bound, we have
\begin{align}
&\text{Pr}\big[E_{(kj)2}\big]\le \sum_{ ({\mathbf w}_{j}, \mathbf l_{j-1})\ne (\mathbf 1, \mathbf 1) }
\text{Pr}\big[E_{1(kj)}(\mathbf w_j, \mathbf l_{j-1},
\mathbf 1)\big]\notag\\
&\overset{(a)}{\le} {\sum_{ ({\mathbf w}_{j}, \mathbf l_{j-1})\ne (\mathbf 1, \mathbf 1) }} 2^{-n(I_{\set
S(\mathbf w_j, \mathbf l_{j-1})}-\delta_{\epsilon_1}(n) ) }\notag\\
&\overset{(b)}{=}\underset{\set S\ne \emptyset}{\sum_{\set S: k\in \cset S}}\;\; \underset{\set S(\mathbf w_j, \mathbf
l_{j-1})=\set S } { \sum_{({\mathbf w}_{j}, \mathbf l_{j-1}):} }2^{-n( I_{\set
S}-\delta_{\epsilon_1}(n) )}\notag\\
&\overset{(c)}= \underset{\set S\ne \emptyset}{\sum_{\set S: k\in \cset S}} \underset{\set M \cup \set Q=\set S} {\sum_{\set M
\subseteq \set S, \set Q \subseteq \set S} } \left( \prod_{i\in \set M} (2^{nR_i} -1 ) \prod_{i\in \set Q} (2^{n\hat R_i} -1
)\right) \notag\\&\quad\quad\quad\quad\quad\quad\quad\quad\cdot 2^{-n( I_{\set S}-\delta_{\epsilon_1}(n))} \notag\\
&< \underset{\set S\ne \emptyset}{\sum_{\set S: k\in \cset S}} \underset{\set M \cup \set Q=\set S} {\sum_{\set M
\subseteq \set S, \set Q \subseteq \set S} } 2^{nR_\set M} 2^{n\hat R_\set Q} 2^{-n( I_{\set S}-\delta_{\epsilon_1}(n)
)}\notag\\
&\overset{(d)}{\le} \underset{\set S\ne \emptyset}{\sum_{\set S: k\in \cset S}} 3^{|\set S|}2^{n\left (R_\set S+\hat R_\set S-
(I_{\set S}-\delta_{\epsilon_1}(n)) \right)}\notag\\
&=\underset{\set S\ne \emptyset}{\sum_{\set S: k\in \cset S}} 2^{n\left [R_\set S-( I_{\set
S}-\hat R_\set S-\frac{{|\set S|}\log_2{3} }{n}- \delta_{\epsilon_1}(n) ) \right]}\label{ErrorBlock4}
\end{align}
where 
\begin{enumerate}[(a)]
\item follows from (\ref{PEISJ}) 
\item follows by collecting the $({\mathbf w}_{j}, \mathbf l_{j-1})\ne (\mathbf 1, \mathbf 1) $ into classes where $\set S=\set
S  ({\mathbf w}_{j}, \mathbf l_{j-1})$
\item follows because there are
\begin{align}
\prod_{i\in \set M} (2^{nR_i} -1 ) \prod_{i\in \set Q} (2^{n\hat R_i} -1 ) \label{sameSet}
\end{align}
different $({\mathbf w}_{j}, \mathbf l_{j-1})\ne (\mathbf 1, \mathbf 1) $ that result in the same $\set M$ and $\set Q$ such that
$\set M\subseteq \set S$, $\set Q\subseteq \set S$ and $\set S=\set M\cup \set Q$
 \item is because for every node $i \in \set S$, we must have one of the following three cases occur: 
\begin{enumerate}[1)]
 \item $i \in \set M$ and $i \notin \set Q$
\item  $i \notin \set M$ and $i \in \set Q$
\item  $i \in \set M$ and $i \in \set Q$
\end{enumerate}
so there are $3^{|\set S|}$ different ways of choosing $\set M$ and $\set Q$.
\end{enumerate}
Since we require $\hat R_k\ge I(\hat Y_k;Y_k|X_k) + \delta_{\epsilon}(n)$, we have
\begin{align}
\label{Motzkin}
I_{\set{S}}-{\hat R_\set S}& \le I_{\mathcal S}-{\sum_{i\in \set S} I(\hat Y_i;Y_i|X_i)}-\delta_{\epsilon}(n)\notag\\
&= I^{\set K}_{\set S}(k)-\delta_{\epsilon}(n).
\end{align}
Combining (\ref{UnionBound}), (\ref{rdbound}), (\ref{ErrorBlock4}) and (\ref{Motzkin}) we find that we
can make $\text{Pr}\big[E_{kj}\big]\rightarrow 0$ as $n\rightarrow \infty$ if
\begin{align}
\label{rate2}
0 \le R_{\set S}&<I^{\set K}_{\mathcal S}(k)
\end{align}
for all subsets $\mathcal S \subset \mathcal K $ such that $k\in \cset S$ and $\mathcal S \ne \emptyset$. Of course, if $I^{\set
K}_{\set S}(k)\le 0$, then we require that $R_{\mathcal S}=0$.

We can split the bounds in (\ref{rate2}) into two classes:
\begin{align}
\text{Class}\; 1: &\;\set S \cap \set {\wt D}_k \ne \emptyset\\
\text{Class}\; 2: & \;\set S \cap \set {\wt D}_k = \emptyset \;\text{or equivalently}\; \set S \subseteq \cset {\wt
D}_k \label{constraintS1}
\end{align}
LNNC requires only the Class 1 bounds. SNNC requires both the Class 1 and Class 2 bounds to guarantee reliable decoding of the
quantization indices $\mathbf l_{j-1}$ for each backward decoding step. With the same argument as in \cite[Sec. IV-C]{Ya02}, we 
can show that the Class 2 bounds can be ignored when determining the best SNNC rates. SNNC with backward decoding thus performs 
as well as SNNC with sliding window decoding and LNNC with joint decoding.

\section{Discussion}
\label{diss}

\subsection{Sliding Window Decoding}
SNNC with sliding window decoding was studied in \cite{Ya02, Ya03} and LNNC \cite{Gamal01} 
achieves the same rates as in \cite{Ya02}. SNNC has extra constraints that turn out to be redundant \cite[Sec. IV-C]{Ya02}, 
\cite[Sec. V-B]{Ya03}. The sliding window decoding in \cite{Ya02} resembles that in \cite{Kramer06} 
where encoding is delayed (or offset) and different decoders are chosen depending on the rate point. The rates achieved by one 
decoder may not give the entire rate region of Theorem~\ref{thm:theorem1}, but the union of achievable rates of all decoders does 
\cite[Theorem 1]{Ya03}. The advantage of sliding window decoding is a small decoding delay of $K+1$ blocks as compared to 
backward decoding that requires $B+K(K-1)$ blocks, where $B\gg K$.

\subsection{Backward Decoding}
SNNC with {\em backward decoding} was studied in \cite{Xie02} for single source networks. For these networks, \cite{Xie02} showed 
that LNNC and SNNC achieve the same rates. Further, for a fixed random coding distribution there is a subset of the relay nodes 
whose messages should be decoded to achieve the best LNNC and SNNC rates. Several other interesting properties of the coding 
scheme were derived. It was also shown in \cite{Pramod01} that SNNC with a layered network analysis \cite{Suhas01} achieves the 
same LNNC rates for single source networks. In \cite{Du01}, SNNC with partial cooperation between the sources was considered for 
multi-source networks.



\subsection{Multihopping}
We compare how the approaches of Theorem~\ref{thm:theorem1} and \cite[Theorem 2.5]{Xie02} reliably convey the last quantization
indices $\mathbf l_B$. Theorem~\ref{thm:theorem1} uses multihopping while Theorem 2.5 in \cite{Xie02} uses a QF-style method with
$M$ extra blocks after block $B$ with the same block length $n$. In these $M$ blocks every node transmits as before except that
the messages are set to a default value. The initialization method in \cite{Xie02} has two disadvantages:
\begin{itemize}
  \item Both $B$ and $M$ must go to infinity to reliably decode $\mathbf l_B$ \cite[Sec.IV-A, Equ. (34)]{Xie02}. The true rate of 
node $k$'s message $w_k$ is
\begin{align}
R^\prime_{k,\text{true}}=\frac{nBR_k}{nB + nM}=\frac{B}{B + M}\cdot R_k \label{badTrueRate}
\end{align}
and we choose $B\gg M$ so that $R^\prime_{k,\text{true}} \rightarrow R_k$ as $B\rightarrow \infty$.
\item {\em Joint} rather than {\em per-block} processing is used.
\end{itemize}
We remark that multihopping may be a better choice for reliably communicating $\mathbf l_B$, because the QF-style approach has a 
large decoding delay due to the large value of $M$ and does not use per-block processing.


\subsection{Choice of Typicality Test}
\begin{figure}[t]
\centering
\subfigure[Both nodes 1 and 2 are sources and
relays]{\psfrag{1}[][][1]{$1$}
\psfrag{2}[][][1]{$2$}
\psfrag{d}[][][1]{$3$}
\includegraphics[width=2.88cm]{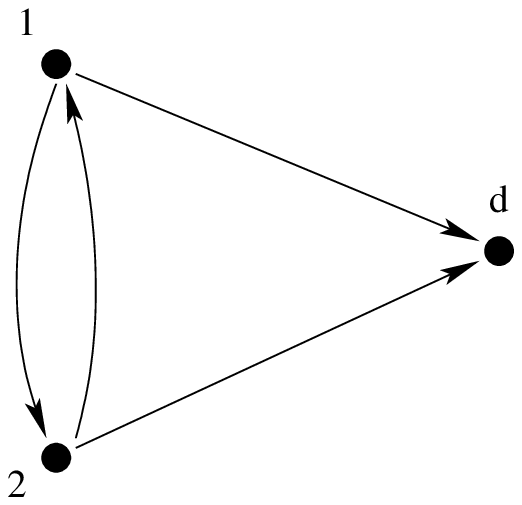}\label{bothRelay}}
\quad\quad\quad
\subfigure[Node 2 acts as a relay for node
1]{\psfrag{1}[][][1]{$1$}
\psfrag{2}[][][1]{$2$}
\psfrag{d}[][][1]{$3$}
\includegraphics[width=2.8cm]{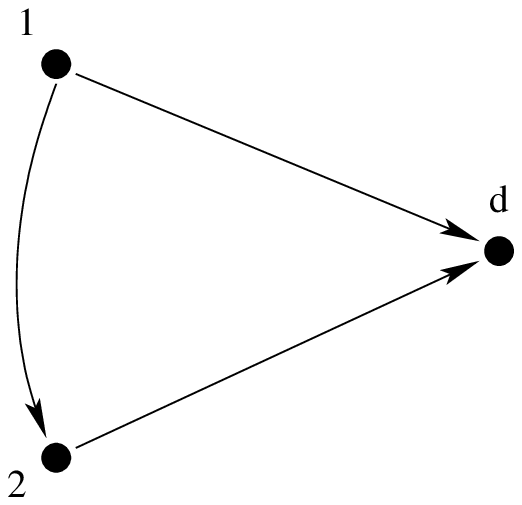}\label{oneRelay}}
\caption{Examples of a three-node network with different rate pairs.}
\label{3node}
\end{figure}

Theorem 1 has a subtle addition to \cite{GH02} and difference to \cite[Theorem 2]{Gamal01} and\cite[Theorem 18.5]{Gamal03}, namely
that in (\ref{Theo1}) each $k\in \set K$ may have a {\em different} set $\set {\wt K}_k$ of nodes satisfying all Class $2$
constraints whose messages and
quantization indices are included in the typicality test. But we can achieve the rates in (\ref{Theo1}) at node $k$ with
SNNC by using backward decoding and treating the signals from the nodes in $\mathcal K \setminus \set {\wt K}_k$ as noise. Hence
we may ignore the Class $2$ constraints in (\ref{constraintS1}) when determining the {\em best} SNNC rates.

\begin{figure}[t]
\centering
\psfrag{R1}[][][1]{$R_1$}
\psfrag{R2}[][][1]{$R_2$}
\psfrag{Point1}[][][0.9]{Point $1$}
\psfrag{Point2}[][][0.9]{Point $ 2$}
\includegraphics[width=4cm]{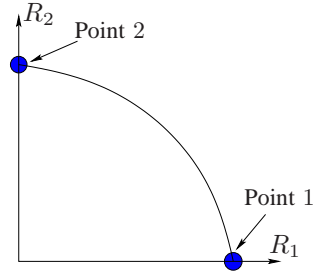}
\caption{Illustration of the achievable rates for the network of Fig.~\ref{oneRelay}.}
\label{rate3node}
\end{figure}

The following example suggests that it may not be surprising that the SNNC and LNNC rate regions are the same. Consider the
network in Fig.~\ref{3node}, where $\set K=\{1,2,3\}$. Suppose both nodes 1 and 2 act as sources as well as relays for each other
in transmitting information to node $3$ (see Fig.~\ref{bothRelay}). Referring to Theorem~\ref{thm:theorem1}, the SNNC
and LNNC bounds are (see Fig.~\ref{rate3node}):
\begin{align}
R_1&<I(X_1;\hat Y_2 Y_3|X_2)-I(\hat Y_1;Y_1|X_1X_2\hat Y_2 Y_3)\label{match1}\\
R_2&<I(X_2;\hat Y_1 Y_3|X_1)-I(\hat Y_2;Y_2|X_1X_2\hat Y_1 Y_3)\label{match2}\\
R_1+R_2&<I(X_1X_2;Y_3)-I(\hat Y_1\hat Y_2;Y_1Y_2|X_1X_2Y_3)\label{match3}
\end{align}
However, suppose now that node 2 has no message ($R_2=0$) and acts as a relay node only (see Fig.~\ref{oneRelay}). Then LNNC does
not have the bound (\ref{match2}) while SNNC has the bound (\ref{match2}) with $R_2=0$ and $\hat Y_1=\emptyset$. We ask
whether (\ref{match2}) reduces the SNNC rate. This is equivalent to asking whether SNNC achieves point $1$ in
Fig.~\ref{rate3node}. It would be strange if there was a discontinuity in the achievable rate region at $R_2=0$.

\subsection{Joint Decoding}
It turns out that SNNC with joint decoding achieves the same rates as in Theorem~\ref{thm:theorem1}. Recently, the authors
of \cite{Buddy01} showed that SNNC with joint decoding fails to achieve the LNNC rates for a {\em specific} choice of SNNC
protocol. However, by multihopping the last quantization indices and then performing joint decoding with the messages and 
remaining quantization bits, SNNC with joint decoding performs as well as SNNC with sliding window or backward decoding, and 
LNNC. This makes sense, since joint decoding should perform at least as well as backward decoding. Details are given in 
Appendix~\ref{appA}.

\subsection{Decoding Subsets of Messages}
From Theorem~\ref{thm:theorem1} we know that if node $k$ decodes messages from nodes in $\set K_k$ and some of the
Class $2$ constraints in (\ref{constraintS1}) are violated, then we should treat the signals from the corresponding nodes as
noise. In this way, we eventually wind up with some $\wt {\set K}_k=\{k\}\cup \wt {\set D}_k \cup \set T_k$, $\set T_k\subseteq
\cset {\wt D}_k\setminus \{k\}$, where all Class 2 constraints are satisfied, i.e., we have
\begin{align}
0\le R_{\set S}<I^{\wt {\set K}_k}_{\set S}(k|T),\;\text{for all} \;\set S\subseteq \set T_k,\; \set S\ne
\emptyset \label{constraintALLS}
\end{align}
and we achieve as good or better rates. In this sense, the sets $\wt {\set K}_k$ are important even for LNNC. These sets seem
difficult to find in large networks because many constraints need to be checked. However, provided that the
sets $\wt {\set K}_k$ are known, we have the following lemma.
\begin{lemma}
For the $K$-node DMN, the rate tuples $(R_{1},\dots,R_{K})$ are achievable if
\begin{align*}
R_{\set{S}}<I^{\wt {\set K}_k}_{\mathcal S}(k|T)
\end{align*}
for all $k\in \set K$, all subsets $\set S \subset \wt {\set K}_k$ with $k\in \cset S$ and $\set S\cap \wt {\set D}_k\ne
\emptyset$, $\wt {\set K}_k=\{k\}\cup \wt {\set D}_k \cup \set T_k$, $\set T_k\subseteq
\cset {\wt D}_k\setminus \{k\}$, where $\set T_k$ satisfies (\ref{constraintALLS}) and for any joint
distribution that factors as (\ref{distribution1}).
\end{lemma}

\begin{IEEEproof}
The proof follows by including the messages from nodes in $\wt {\set K}_k$ satisfying (\ref{constraintALLS}) in the
typicality test at every destination $k$ in Theorem~\ref{thm:theorem1}.
\end{IEEEproof}

\subsection{Optimal Decodable Sets}
\label{optimalSet}
SNNC was studied for relay networks in \cite{Xie02}. For such networks there is one message at node $1$ that is destined for node
$K$. We thus have $\set {\wt D}_K=\{1\}$ and $\cset {\wt D}_K\setminus \{K\}=\{2,\dots, K-1\}$. The authors
of \cite{Xie02} showed that for a given random coding distribution 
\begin{align}
P(t)P(x_1|t)\prod_{k\in \cset{\wt D}_K}P(x_k|t)P(\hat y_k|y_k,x_k, t) 
\end{align}
there exists a {\em unique largest optimal decodable set} $\set {T}^\ast$, $\set {T}^\ast \subseteq \cset {\wt D}_K\setminus
\{K\}$, of the relay nodes that provides the same best achievable rates for both SNNC and LNNC \cite[Theorem 2.8]{Xie02}. We now
show that the concept of {\em optimal decodable set} extends naturally to {\em multi-source networks}.
\begin{lemma}
\label{lem:lemma2}
For a $K$-node memoryless network with a fixed random coding distribution 
\begin{align}
P(t)\prod^K_{k=1}P(x_k|t)P(\hat y_k|y_k,x_k, t)
\end{align}
there exists for each node $k$ a {\em {unique largest}} set $\set T^\ast_k$ among all subsets $\set T_k\subseteq \cset {\wt D}_k
\setminus \{k\}$ satisfying (\ref{constraintALLS}). The messages of the nodes in $\set T^\ast_k$ should be included in the
typicality test to provide the best achievable rates.
\end{lemma}

\begin{IEEEproof}
We prove Lemma~\ref{lem:lemma2} without a time-sharing random variable $T$. The proof with $T$ is similar. We show that $\set
T^\ast_k$ is unique by showing that the union of any two sets $\set T^{1}_k$ and $\set T^{2}_k$ satisfying all constraints also
satisfies all constraints and provides as good or better rates. Continuing taking the union, we eventually reach a unique largest
set $\set T^\ast_k$ that satisfies all constraints and gives the best rates.

Partition the subsets $\set T_k\subseteq \cset {\wt D}_k\setminus \{k\}$ into two classes:

Class $1$: $R_{\set S}<I^{\set K_k}_{\set S}(k)$ for all $\set S \subseteq \set T_k$;

Class $2$: There exists one $\set S \subseteq \set T_k$ such that $R_{\set S}\ge I^{\set K_k}_{\set S}(k)$.

We may ignore the $\set T_k$ in Class $2$ because the proof of Theorem~\ref{thm:theorem1} shows that we can treat the signals of
nodes associated with violated constraints as noise and achieve as
good or better rates. Hence, we focus on $\set T_k$ in Class $1$.

Suppose $\set T^1_{k}$ and $\set T^2_{k}$ are in Class $1$ and let
$\set T^3_{k}=\set T^1_{k} \cup \set T^2_{k}$. We define 
\begin{align}
\set {\wt K}^1_{k}&=\{k\}\cup \set {\widetilde D}_k  \cup \set T^1_{k}\\
\set {\wt K}^2_{k}&=\{k\}\cup \set {\widetilde D}_k  \cup \set T^2_{k}\\
\set {\wt K}^3_{k}&=\{k\}\cup \set {\widetilde D}_k  \cup \set T^3_{k}.
\end{align}
Further, for every $\set S\subseteq \set {\wt K}^{3}_{k}$, define $\set S_1=\set S \cap \set {\wt K}^{1}_{k}$ and
$\set S_2=\set S \cap (\set {\wt K}^{3}_{k} \setminus \set S_1)$. We have $\set S_1\subseteq \set {\wt K}^{1}_{k} $,
$\set S_2\subseteq \set {\wt K}^{2}_{k}$, $\set S_1\cup \set S_2=\set S$ and $\set S_1\cap \set S_2=\emptyset$.
We further have
\begin{align}
&R_{\set S}\overset{(a)}{=}R_{\set S_1}+R_{\set S_2}\notag\\
&\overset{(b)}{<}I^{\set {\wt K}^{1}_{k}}_{\set S_1}(k)+I^{\set {\wt K}^{2}_{k}}_{\set S_2}(k)\notag\\
&\overset{(c)}{=} I(X_{\set S_1};\hat Y_{\set {\wt K}^{1}_{k} \setminus \set S_1} Y_k|X_{\set {\wt K}^{1}_{k} \setminus \set
S_1})-I(\hat Y_{\set S_1};Y_{\set S_1}|X_{\set {\wt K}^{1}_{k}} \hat Y_{\set {\wt K}^{1}_{k}\setminus \set S_1}Y_k)\notag\\
&\;\;+  I(X_{\set S_2};\hat Y_{\set {\wt K}^{2}_{k} \setminus \set S_2} Y_k|X_{\set {\wt K}^{2}_{k} \setminus \set S_2})-I(\hat
Y_{\set S_2};Y_{\set S_2}|X_{\set {\wt K}^{2}_{k}} \hat Y_{\set {\wt K}^{2}_{k}\setminus \set S_2}Y_k)\notag\\
&\overset{(d)}{\le} I(X_{\set S_1}; \hat Y_{\set {\wt K}^{3}_{k}\setminus \set S }Y_k|X_{\set {\wt K}^{3}_{k}\setminus \set S
})-I(\hat Y_{\set S_1};Y_{\set S_1}|X_{\set {\wt K}^{1}_{k}} \hat Y_{\set {\wt K}^{1}_{k}\setminus \set S_1}Y_k)\notag\\
&\;\;+I(X_{\set S_2};\hat Y_{\set {\wt K}^{3}_{k} \setminus \set S} Y_k|X_{\set {\wt K}^{3}_{k} \setminus \set S_2})-I(\hat
Y_{\set S_2};Y_{\set S_2}|X_{\set {\wt K}^{2}_{k}} \hat Y_{\set {\wt K}^{2}_{k}\setminus \set S_2}Y_k)\notag\\
&\overset{(e)}{\le} I(X_{\set S_1}; \hat Y_{\set {\wt K}^{3}_{k}\setminus \set S }Y_k|X_{\set {\wt K}^{3}_{k}\setminus \set
S })-I(\hat Y_{\set S_1};Y_{\set S_1}|X_{\set {\wt K}^{3}_{k} } \hat Y_{\set {\wt K}^{3}_{k}\setminus \set S}Y_k)\notag\\
&\;\;+ I(X_{\set S_2};\hat Y_{\set {\wt K}^{3}_{k} \setminus \set S} Y_k|X_{\set {\wt K}^{3}_{k} \setminus \set S_2})-I(\hat
Y_{\set S_2};Y_{\set S_2}|X_{\set {\wt K}^{3}_{k}} \hat Y_{\set {\wt K}^{3}_{k}\setminus \set S_2}Y_k)\notag\\
&\overset{(f)}{=}I(X_\set S; \hat Y_{\set {\wt K}^{3}_{k}\setminus \set S }Y_k|X_{\set {\wt K}^{3}_{k}\setminus \set S })-I(\hat
Y_{\set S};Y_{\set S}|X_{\set {\wt K}^{3}_{k}} \hat Y_{\set {\wt K}^{3}_{k}\setminus \set S}Y_k)\notag\\
&\overset{(g)}{=}I^{\set {\wt K}^{3}_{k}}_{\set S}(k)\label{UnionBetter}
\end{align}
where
\begin{enumerate}[(a)]
\item follows from the definition of $\set S_1 $ and $\set S_2 $
\item follows because both $\set T^1_k$ and $\set T^2_k$ are in Class $1$
\item follows from the definition (\ref{RateNoTime})
\item follows because all $X_k$ are independent and conditioning does not increase entropy 
\item follows because conditioning does not increase entropy and by the Markov chains
\begin{align}
&X_{\set {\wt K}^{3}_{k} \setminus \set S_2}\hat Y_{\set {\wt K}^{3}_{k} \setminus \set S_2} Y_k - Y_{\set S_2}X_{\set S_2}-\hat
Y_{\set S_2}\label{chain3}\\&X_{\set {\wt K}^{3}_{k} \setminus \set S_1}\hat Y_{\set {\wt K}^{3}_{k} \setminus \set S} Y_k
-Y_{\set S_1}X_{\set S_1}-\hat Y_{\set S_1}\label{chain4}
\end{align}
\item follows from the chain rule for mutual information and the Markov chains (\ref{chain3}) and
(\ref{chain4})
\item follows from the definition (\ref{RateNoTime}).
\end{enumerate}
The bound (\ref{UnionBetter}) shows that $\set T^3_k$ is also in Class $1$. Moreover, by (\ref{UnionBetter}) if $k$ includes the
messages of nodes in $\set {\wt K}^{3}_{k}$ in the typicality test, then the rates are as good or better than those achieved by
including the messages of nodes in $\set {\wt K}^{1}_{k}$ or $\set {\wt K}^{2}_{k}$ in the typicality test. Taking the union of
all $\set T_k$ in Class $1$, we obtain the {\em unique largest} set $\set T^\ast_k$ that gives the best achievable rates.
\end{IEEEproof}

\begin{remark}
\label{rem:remark5}
There are currently no efficient algorithms for finding an optimal decodable set. Such algorithms would be useful for
applications with time-varying channels. 
\end{remark}

\section{SNNC with a DF option}
\label{mixed}

\begin{table*}
\begin{center}
\begin{tabular}{r|c c c c c c}
  \hline
  Block   &       1            &        2           &       $\cdots$   &            $B$        & $B+1$ &          $B+2 \cdots
B+4$\\
\hline
  $X_1$ & $\mathbf x_{11}(w_{1},1)$ & $\mathbf x_{12}(w_{2}, w_{1})$    &  $\cdots$     &  $\mathbf x_{1B}(w_{B}, w_{B-1})$
& $\mathbf x_{1(B+1)}(1, w_{B})$ & \\
  $X_2$ & $\mathbf x_{21}(1)$     & $\mathbf x_{22}(w_1)$ &  $\cdots$   &  $\mathbf x_{2B}(w_{(B-1)})$ & $\mathbf
x_{2(B+1)}(w_{B})$ & Multihop $l_{B+1}$ \\
$X_3$ & $\mathbf x_{31}(1)$   & $\mathbf x_{32}(l_1)$ & $\cdots$   &  $\mathbf x_{3B}(l_{B-1})$ & $\mathbf x_{3(B+1)}(l_{B})$& to
node $4$ in $3n^\prime$\\ 
$\hat Y_3$  & $\mathbf {\hat y}_{31}(l_{1}|1)$ & $\mathbf {\hat y}_{32}(l_{2}|l_1)$ &$\cdots$   & $\mathbf {\hat
y}_{3B}(l_{B}|l_{B-1})$ & $\mathbf {\hat y}_{3(B+1)}(l_{B+1}|l_{B})$& channel uses\\
\hline
\end{tabular}
\end{center}
\caption{Coding scheme for the two-relay channel with block Markov coding at the source.}
\label{tworelay}
\end{table*}
One of the main advantages of SNNC is that the relays can switch between QF (or CF) and DF depending on the channel conditions. If
the channel conditions happen to be good, then the natural choice is DF which removes the noise at the relays. This not possible
with LNNC due to the high rate of the long message. On the other hand, if a relay happens to experience a deep fade, then this 
relay should use QF (or CF).

In the following, we show how mixed strategies called SNNC-DF work for the multiple-relay channel. These mixed strategies are
similar to those in \cite[Theorem 4]{Kramer04}. However, in \cite{Kramer04} the relays use CF with a prescribed binning rate to
enable {\em step-by-step} decoding (CF-S) instead of QF. In Section~\ref{Gauss}
we give numerical examples to show that SNNC-DF can outperform DF, CF-S and LNNC.

As in \cite{Kramer04}, we partition the relays $\set T=\{2,\dots,K-1\}$ into two sets
\begin{align*}
&\set T_1=\{k:\; 2 \le k \le K_1 \}\\
&\set T_2=\set T \setminus \set T_1
\end{align*}
where $1\le  K_1 \le K-1$. The relays in $\set T_1$ use DF while the relays in $\set T_2$ use QF. 
Let $\pi(\cdot)$ be a permutation on $\{1,\dots,K\}$ with $\pi(1)=1$ and $\pi(K)=K$ and let
$\pi(j:k)=\{\pi(j),\pi(j+1),\dots,\pi(k)\}$, $1\le j\le k \le K$. Define $\set T_{i(\pi)}=\{ \pi(k), k \in \set T_i\},\;
i=1,2$. We have the following theorem. 

\begin{theorem}
\label{thm:theorem2} 
SNNC-DF achieves the rates satisfying
\begin{align}
&R_\text{SNNC-DF}<\underset{\pi(\cdot)}{\text{max}}\:\underset{K_1}{\text{max}}\:\text{min}\;\notag\\&\left\{
\underset{1\le k\le K_1-1}{\text{min}} I\left(X_{\pi(1:k)};Y_{\pi(k+1)}|X_{\pi(k+1:K_1)}\right),\right.\notag\\
&\left.I(X_1X_{\set T_{1(\pi)}} X_{\mathcal S}; \hat Y_{\cset S}Y_K|X_{\cset S})-I(\hat Y_{\mathcal S};
Y_\mathcal S|X_1X_{\set T}\hat Y_{\cset S} Y_K)\right\}\label{rateDQF}
\end{align}
for all $\set S\subseteq \set T_{2(\pi)}$, where $\cset S$ is the complement of $\mathcal S$ in $\mathcal T_{2(\pi)}$, and
where the joint distribution factors as
\begin{align}
P(x_1x_{\set T_{1(\pi) }})\cdot & \Big[\prod_{k\in \mathcal T_{2(\pi)}}P(x_k)P(\hat y_k|y_k,x_k) \Big] \notag\\ 
&\cdot  P(y_2,\dots,y_K|x_1,\dots,x_{K-1}).
\end{align}
 
\end{theorem}

\begin{remark}
As usual, we may add a time-sharing random variable to improve rates.
\end{remark}

\begin{IEEEproof}[Proof Sketch]
For a given permutation $\pi(\cdot)$ and $K_1$, the first mutual information term in (\ref{rateDQF}) describes the DF
bounds \cite[Theorem 1]{Kramer04} (see also \cite[Theorem 3.1]{Xie01}). The second mutual information term in (\ref{rateDQF})
describes the SNNC bounds. Using a similar analysis as for Theorem~\ref{thm:theorem1} and by treating $(X_1X_{\set T_{1(\pi)}})$
as the \lq\lq new\rq\rq\:source signal at the destination, we have the SNNC bounds
\begin{align}
R_\text{SNNC-DF}&<I(X_1X_{\set T_{1(\pi)}} X_{\mathcal S}; \hat Y_{\cset S}Y_K|X_{\cset S})\notag\\&\quad\quad\quad\quad-I(\hat
Y_{\mathcal S};
Y_\mathcal S|X_1X_{\mathcal T}\hat Y_{\cset S} Y_K)\\
0&\le I(X_{\mathcal S}; \hat Y_{\cset S}Y_K|X_1X_{\mathcal T_{1(\pi)}}X_{\cset S})\notag
\\&\quad\quad\quad\quad-I(\hat Y_{\mathcal S}; Y_\mathcal S|X_1X_{\mathcal T}\hat Y_{\cset S} Y_K)  \label{consDQF}
\end{align} for all $\mathcal S\subseteq \mathcal T_{2(\pi)}$.

The same argument used to prove Theorem~\ref{thm:theorem1} shows that if any of the
constraints (\ref{consDQF}) is violated, then we get rate bounds that can be achieved with SNNC-DF by
treating the signals from the corresponding relay nodes as noise. Thus we may ignore the constraints (\ref{consDQF}).
\end{IEEEproof}
\begin{example}
\label{exp:example4}
Consider $K=4$ and $K_1=2$. There are two possible permutations $\pi_1(1:4)=\{1,2,3,4\}$ and
$\pi_2(1:4)=\{1,3,2,4\}$. For $\pi_1(1:4)=\{1,2,3,4\}$, Theorem~\ref{thm:theorem2} states that SNNC-DF achieves any rate
up to \begin{align}
R_\text{SNNC-DF}&={\text{min}}\;\Big\{ I(X_1;Y_2|X_2), I(X_1X_2; \hat Y_3Y_4|X_3),\notag\\
&I(X_1X_2X_3;Y_4)-I(\hat Y_{3}; Y_3|X_1X_2X_3Y_4)\Big\}\label{MarkovBack}
\end{align}
where the joint distribution factors as
\begin{align}
\label{distributionMarkov}
P(x_1,x_2)P(x_3)P(\hat y_3|y_3,x_3)\cdot P(y_2,y_3,y_4|x_1,x_2,x_3).
\end{align}
The corresponding coding scheme is given in Table~\ref{tworelay}. 
\end{example}

If relay node 2 uses DF while relay node 3 uses CF-S, then by \cite[Theorem 4]{Kramer04} with $U_2=0$, any rate up to
\begin{align}
R_\text{[CF-S]-DF}<{\text{min}}\;\left\{ I(X_1;Y_2|X_2), I(X_1X_2; \hat Y_3Y_4|X_3)\right\}\label{DCF}
\end{align}
can be achieved, subject to
\begin{align}
\label{CDCF}
I(\hat Y_3;Y_3|X_3Y_4) \le I(X_3;Y_4) 
\end{align}
and the joint distribution factors as (\ref{distributionMarkov}). It turns out that $R_\text{[CF-S]-DF}$ in
(\ref{DCF})-(\ref{CDCF}) is the same as $R_\text{SNNC-DF}$ (\ref{MarkovBack}), since LNNC and SNNC do not improve the CF-S 
rate for one relay \cite{GH01}. But $R_\text{SNNC-DF}$ is better than $R_\text{[CF-S]-DF}$ in general.

\begin{remark}
For rapidly changing channels it is advantageous to use independent inputs so all nodes can use the same encoder for all channel 
states. If $X_1$ and $X_2$ in the above example are independent, there is no need to use block Markov coding (BMC). However, 
we need to use two backward (or sliding window) decoders to recover the rates (\ref{MarkovBack}). See Appendix \ref{appB}.
\end{remark}

\begin{remark}
 \label{rem:remark6}
How to perform DF for multiple sources is not obvious. Consider again the three node network in Fig.~\ref{3node}, 
but now every node wishes to send a message to the other two nodes. How should one set up cooperation if all nodes may use DF?
Such questions are worth addressing, since their answers will give insight
on how to incorporate mixed strategies to boost system performance.
\end{remark}

\section{Gaussian Networks}
\label{Gauss}
We next consider additive white Gaussian noise (AWGN) networks. We use $X\sim \set C\set N(\mu,\sigma^2) $ to denote a circularly
symmetric complex Gaussian random variable $X$ with mean $\mu$ and variance $\sigma^2$. Let $Z^{K}=Z_1 Z_2\dots Z_K$ be a
noise string whose symbols are i.i.d. and $Z_k \sim \set C\set N(0,1) $ for all $k$. The channel output at node $k$ is
\begin{align}
Y_k=\left[\underset{j\ne k}{\sum^K_{j=1}}{G_{jk}}X_j\right] + Z_k
\end{align}
where the channel gain is
\begin{align}
G_{jk}=\frac{H_{jk}}{\sqrt{d^\alpha_{jk}}}
\end{align}
and $d_{jk}$ is the distance between nodes $j$ and $k$, $\alpha$ is a path-loss exponent and $H_{jk}$ is a complex fading
random variable.

We consider two kinds of fading:
\begin{itemize}
\item No fading: $H_{jk}$ is a constant and known at all nodes. We set $H_{jk}=1$ for all $j,k\in \set K$.

\item Rayleigh fading: we have $H_{jk}\sim \set C\set N(0,1)$. We assume that a destination node $k$ knows
$G_{jk}$ for all $j, k\in \set K$ and a relay node $k$ knows $G_{jk}$ for all $j \in \set K$ and knows the statistics of all other
$G_{jl}, j,l \in \set K$. We focus on slow fading, i.e., all $G_{jk}$ remain unchanged once chosen.
\end{itemize}

We avoid issues of power control by imposing a per-symbol power constraint $\text{E}[|X_k|^2]\le P_k$. We choose the inputs to be
Gaussian, i.e., $X_k\sim \set C\set N(0, P_k)$, $k\in \set K$. 

In the following we give numerical examples for four different channels
\begin{itemize}
 \item the relay channel;
\item the two-relay channel;
\item the multiple access relay channel (MARC);
\item the two-way relay channel (TWRC).
\end{itemize}
We evaluate the performance for no fading in terms of achievable rates (in bits per channel use) and for Rayleigh fading in
terms of outage probability \cite{Ozarow01} for a target rate $R_\text{tar}$.

Relay node $k$ chooses 
\begin{align}
\hat Y_k=Y_k+\hat Z_k
\end{align}
where $\hat Z_k \sim \set C\mathcal N(0,\hat
\sigma^2_k)$. For the no fading case, relay node $k$ numerically calculates the optimal $\hat \sigma^2_k$ for
CF-S and SNNC, and the optimal binning rate $R_{k(\text{bin})}$ for CF-S, in order to maximize the rates. For DF, the source and
relay nodes numerically calculate the power allocation for superposition coding that maximizes the rates. For the Rayleigh
fading case, relay node $k$ knows only the $G_{jk}$, $j \in \set K$, but it can calculate the optimal $\hat \sigma^2_k$ and
$R_{k(\text{bin})}$ based on the statistics of $G_{jl}$, for all $j,l \in \set K$ so as to minimize the outage probability. For
DF, the fraction of power that the source and relay nodes allocate for cooperation is calculated numerically based on the
statistics of $G_{jk}$, for all $j, k\in \set K$, to minimize the outage probability. Details of the derivations are given in
Appendix \ref{appC}.

\subsection{Relay Channels}
\begin{figure}[t!]
\centering
\psfrag{1}[][][1]{$1$}
\psfrag{2}[][][1]{$2$}
\psfrag{3}[][][1]{$3$}
\psfrag{d12}[][][1]{$d_{12}$}
\psfrag{d23}[][][1]{$d_{23}=1-d_{12}$}
\psfrag{d13}[][][1]{$d_{13}=1$}
{\includegraphics[width=5.5cm]{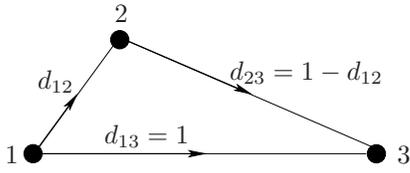}}
\caption{A relay channel.}
\label{relayChannel}
\end{figure}

The Gaussian relay channel (Fig.~\ref{relayChannel}) has
\begin{align}
Y_2&=G_{12} X_1 + Z_2\\
Y_3&=G_{13} X_1+G_{23} X_2+Z_3
\end{align}
and source node $1$ has a message destined for node $3$.

\begin{figure}[t!]
\centering
{\includegraphics[width=8.8cm]{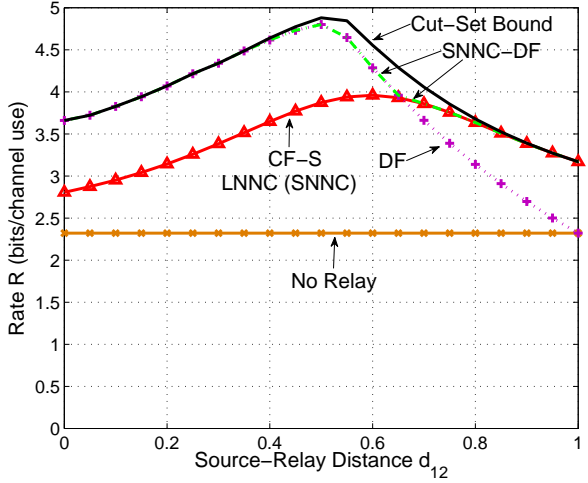}}
\caption{Achievable rates $R$ (in bits per channel use) for a relay channel with no fading.}
\label{rateOneRelay}
\end{figure}

\subsubsection{No Fading}
Fig.~\ref{relayChannel} depicts the geometry and Fig.~\ref{rateOneRelay} depicts the achievable rates as a
function of $d_{12}$ for $P_1=4, P_2=2$ and $\alpha=3$. DF achieves rates close to capacity when the relay is close to the
source while CF-S dominates as the relay moves towards the destination. For the relay channel, CF-S performs as well as SNNC
(LNNC). SNNC-DF unifies the advantages of both SNNC and DF and achieves the best rates for all relay positions.

\subsubsection{Slow Rayleigh fading}
\begin{figure}[t!]
\centering
{\includegraphics[width=8.8cm]{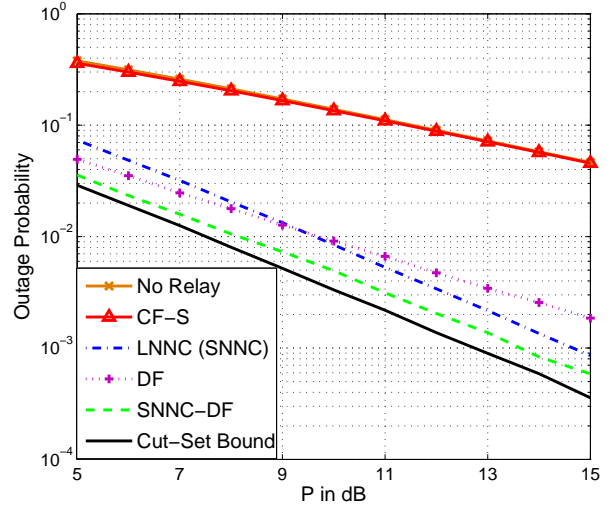}}
\caption{Outage probabilities for a relay channel with Rayleigh fading.}
\label{OutOneRelay}
\end{figure}

Fig.~\ref{OutOneRelay} depicts the outage probabilities with $R_\text{tar}=2$, $P_1=2P, P_2=P$, $d_{12}=0.3,$ $d_{23}=0.7$,
$d_{13}=1$ and $\alpha=3$.

Over the entire power range CF-S gives the worst outage probability. This is because CF-S requires a reliable relay-destination
link so that both the bin and quantization indices can be recovered. Both DF and SNNC improve on CF-S. DF performs better at low
power while SNNC is better at high power. SNNC-DF has the relay decode if possible and perform QF otherwise, and
gains $1$ dB over SNNC and DF.

\subsection{Two-Relay Channels}
The Gaussian two-relay channel (Fig.~\ref{TRC}) has
\begin{align}
Y_2&=G_{12}X_1+G_{32}X_3+Z_2\\
Y_3&=G_{13}X_1+G_{23}X_2+Z_3\\
Y_4&=G_{14}X_1+G_{24}X_2+G_{34}X_3+Z_4
\end{align}
where the relay nodes $2$ and $3$ help node $1$ transmit a message to node $4$.

\subsubsection{No Fading}

\begin{figure}[t]
\centering
\psfrag{1}[][][1]{$1$}
\psfrag{2}[][][1]{$2$}
\psfrag{3}[][][1]{$3$}
\psfrag{4}[][][1]{$4$}
\psfrag{d12}[][][1]{$d_{12}=0.2$}
\psfrag{d13}[][][1]{$d_{13}=0.8$}
\psfrag{d14}[][][1]{$d_{14}=1$}
\psfrag{d23}[][][1]{$d_{23}=0.75$}
\psfrag{d24}[][][1]{$d_{24}=0.8$}
\psfrag{d32}[][][1]{$d_{32}=0.75$}
\psfrag{d34}[][][1]{$d_{34}=0.2$}
{\includegraphics[width=8cm]{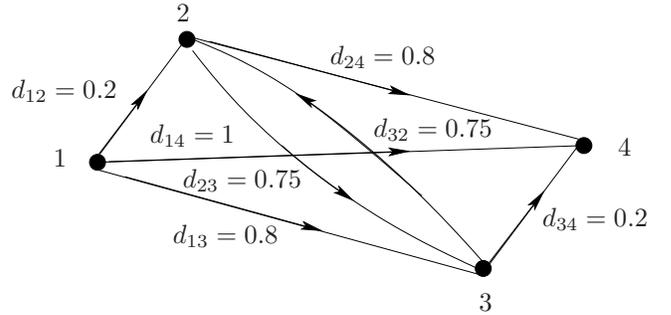}}
\caption{A two-relay channel.}
\label{TRC}
\end{figure}
\begin{figure}[t]
\centering
\includegraphics[width=8.8cm]{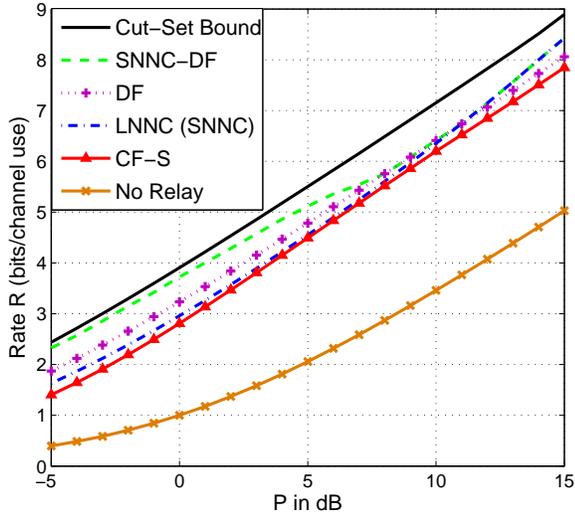}
\caption{Achievable rates R (in bits per channel use) for a TRC with no fading.}
\label{TwoRelay}
\end{figure}

Fig.~\ref{TRC} depicts the geometry and Fig.~\ref{TwoRelay} depicts the achievable rates for $P_1=P_2=P_3=P$ and $\alpha=3$. 
The CF-S rates are the  lowest over the entire power range. As expected, SNNC improves on CF-S. DF performs better than SNNC at
low power but worse at high power. SNNC-DF achieves the best rates and exhibits reasonable rate and power gains over  SNNC and
DF for $P=-5\;\text{dB}$ to $5 \;\text{dB}$. The gains are because in this power range SNNC-DF has relay 2 performing DF and relay
3 performing QF.

\subsubsection{Slow Rayleigh Fading}
\begin{figure}[t!]
\centering
\includegraphics[width=8.8cm]{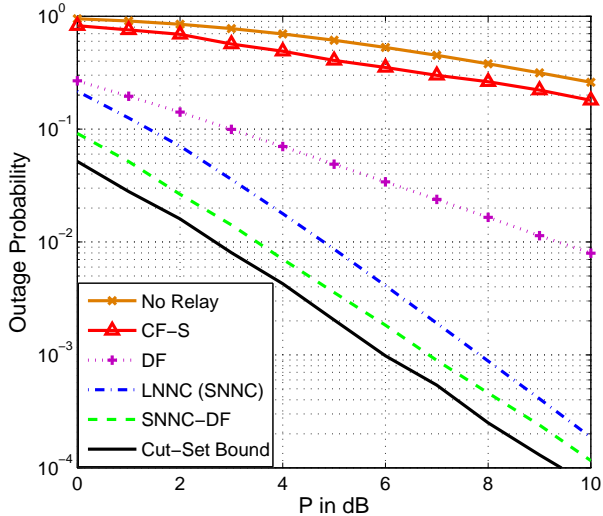}
\caption{ Outage probabilities for a TRC with Rayleigh fading.}
\label{TwoRelayOutage}
\end{figure}

Fig.~\ref{TwoRelayOutage} depicts the outage probabilities with $R_\text{tar}=2$, $P_1=P_2=P_3=P$, the geometry of Fig.~\ref{TRC}
and $\alpha=3$.
CF-S gives the worst performance over the entire power range. This is because CF-S requires a reliable relay-destination link for
{\em both} relays so that the bin and quantization indices for both relays can be decoded. DF provides better outage
probabilities than CF-S but is worse than SNNC or LNNC, since it requires reliable decoding at both relays.  SNNC-DF has the two
relays decode if possible and perform QF otherwise and gains about $1$ dB over LNNC (SNNC). In general, we expect larger gains of
SNNC-DF over LNNC for networks with more relays.

\subsection{Multiple Access Relay Channels}
The Gaussian MARC (Fig.~\ref{MARC}) has
\begin{align}
Y_3&=G_{13}X_1+G_{23}X_2 +Z_3\\
Y_4&=G_{14}X_1+G_{24}X_2+G_{34}X_3+Z_4
\end{align}
and nodes 1 and 2 have messages destined for node 4.

\subsubsection{No Fading}
\begin{figure}[t]
\centering
\psfrag{1}[][][1]{$1$}
\psfrag{2}[][][1]{$2$}
\psfrag{3}[][][1]{$3$}
\psfrag{4}[][][1]{$4$}
\psfrag{d13}[][][1]{$d_{13}=0.75$}
\psfrag{d14}[][][1]{$d_{14}=1$}
\psfrag{d23}[][][1]{$d_{23}=0.55$}
\psfrag{d24}[][][1]{$d_{24}=1$}
\psfrag{d34}[][][1]{$d_{34}=0.45$}
{\includegraphics[width=6.5cm]{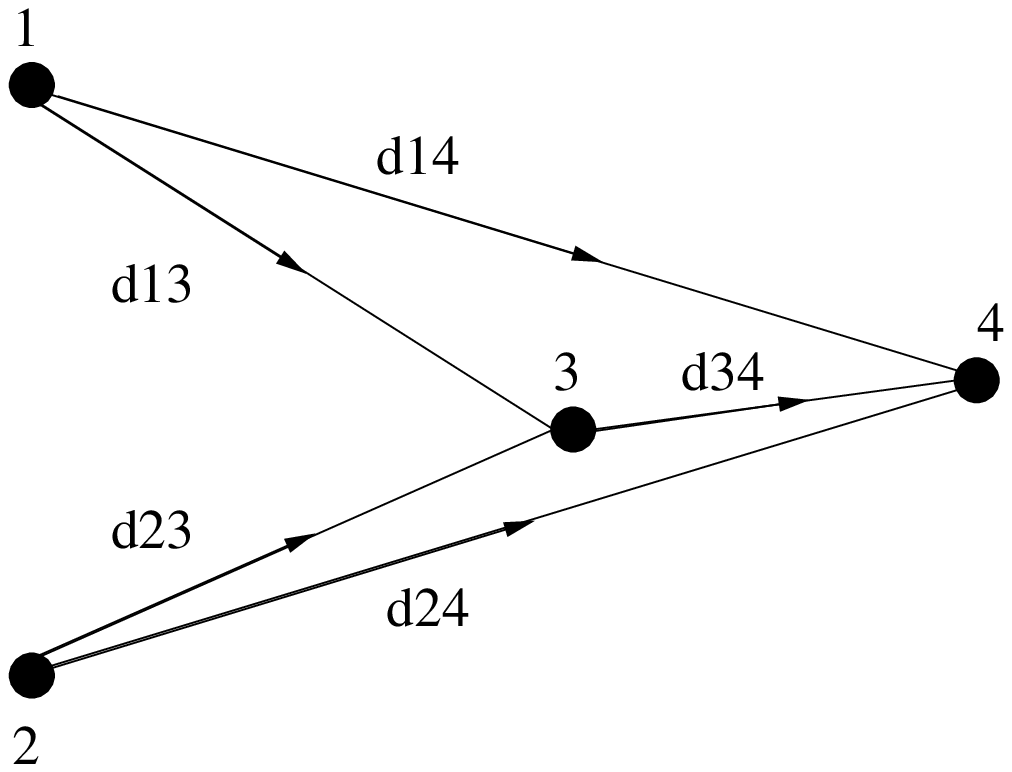}}
\caption{A MARC.}
\label{MARC}
\end{figure}

\begin{figure*}[t]
\centering
\includegraphics[width=12cm]{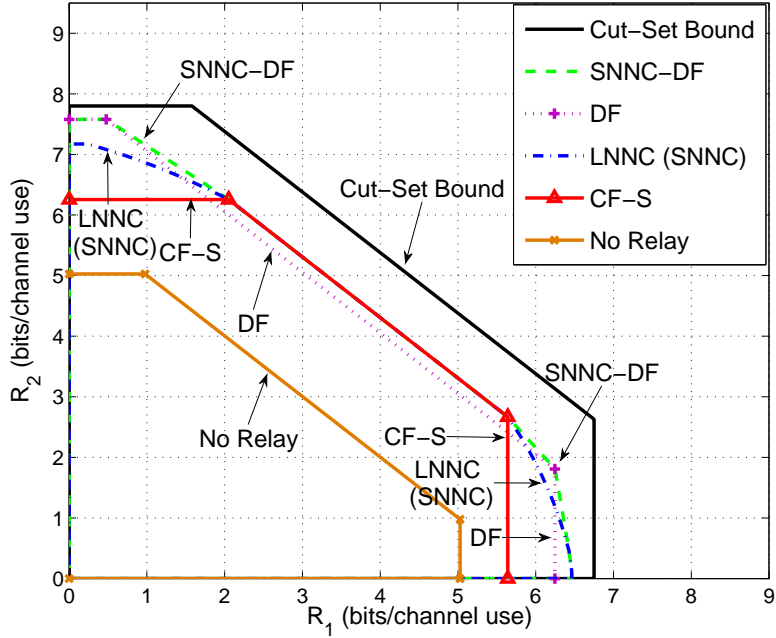}
\caption{Achievable rate regions for a MARC with no fading.}
\label{MARCrate}
\end{figure*}

Fig.~\ref{MARC} depicts the geometry and Fig.~\ref{MARCrate} depicts the achievable rate regions for $P_1=P_2=P_3=P$, $P=15$ dB
and $\alpha=3$. The SNNC rate region includes the CF-S rate region. Through
time-sharing, the SNNC-DF region is the convex hull of the union of DF and SNNC regions. 
SNNC-DF again improves on LNNC (or SNNC) and DF.

\subsubsection{Slow Rayleigh Fading}

\begin{figure}[t]
\centering
\includegraphics[width=8.9cm]{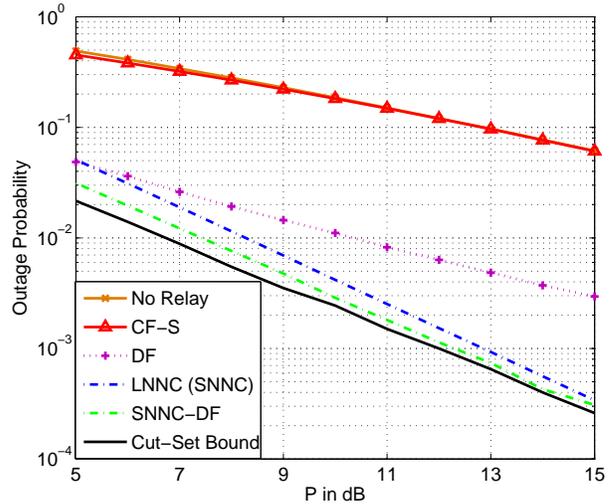}
\caption{Outage probabilities for a MARC with Rayleigh fading.}
\label{MARCOutage}
\end{figure}

Fig.~\ref{MARCOutage} depicts the outage probabilities with $R_\text{tar1}=R_\text{tar2}=1$, $P_1=P_2=P_3=P$,
$d_{13}=0.3$, $d_{23}=0.4$, $d_{14}=d_{24}=1$, $d_{34}=0.6$ and $\alpha=3$.
CF-S has the worst outage probability because it requires a reliable relay-destination link to decode the bin and quantization
indices. DF has better outage probability than CF-S, while LNNC (or SNNC) improves on DF over the entire power range.
SNNC-DF has the relay perform DF or QF depending on channel quality and gains $1$ dB at low power and $0.5$ dB at high
power over SNNC.

\begin{remark}
The gain of SNNC-DF over SNNC is not very large at high power. This is because the MARC has one relay only. For networks with more
relays we expect larger gains from SNNC-DF.
\end{remark}

\subsection{Two-Way Relay Channels}
\begin{figure}[t]
\centering
\psfrag{1}[][][1]{$1$}
\psfrag{2}[][][1]{$2$}
\psfrag{3}[][][1]{$3$}
\psfrag{d12}[][][1]{$d_{12}=1$}
\psfrag{d13}[][][1]{$d_{13}=0.25$}
\psfrag{d23}[][][1]{$d_{23}=0.75$}
{\includegraphics[width=5.8cm]{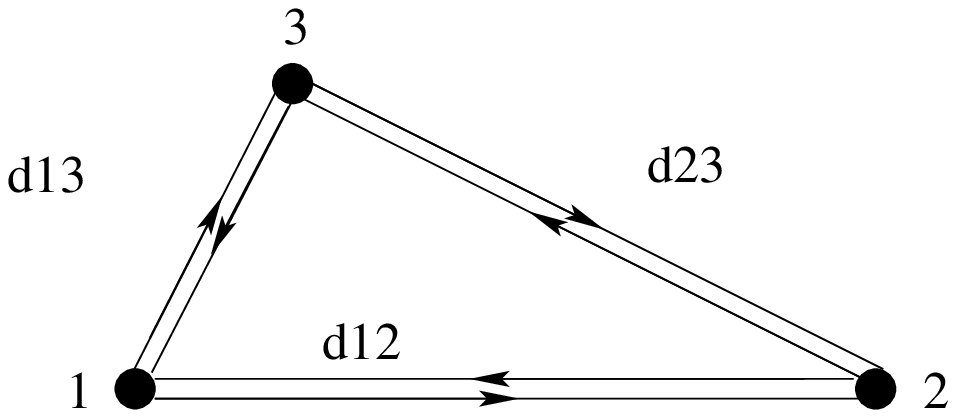}}
\caption{A TWRC.}
\label{TWRC}
\end{figure}

The Gaussian TWRC  (Fig.~\ref{TWRC}) has
\begin{align}
Y_1&=G_{21}X_2+G_{31}X_3+Z_1\\
Y_2&=G_{12}X_1+G_{32}X_3+Z_2\\
Y_3&=G_{13}X_1+G_{23}X_2+Z_3
\end{align}
where nodes 1 and 2 exchange messages with the help of relay node 3.

\subsubsection{No Fading}
\begin{figure}[t]
\centering
\includegraphics[width=8.8cm]{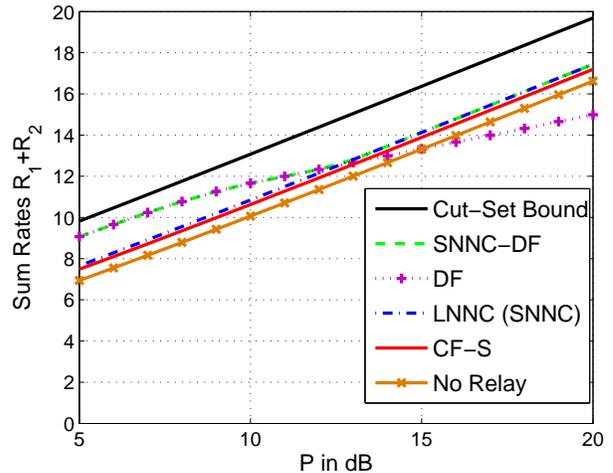}
\caption{Achievable sum rates (in bits per channel use) for a TWRC with no fading.}
\label{rateTWRC}
\end{figure}

Fig.~\ref{TWRC} depicts the geometry and Fig.~\ref{rateTWRC} depicts the achievable sum rates for $P_1=5P, P_2=2P, P_3=P$ and
$\alpha=3$. DF gives the best rates at low power while SNNC provides better rates at high power. The CF-S rates are slightly
lower than the SNNC rates over the entire power range. SNNC-DF combines the advantages of SNNC and DF and achieves the best
rates throughout. 
\subsubsection{Slow Rayleigh Fading}
\begin{figure}[t!]
\centering
{\includegraphics[width=8.8cm]{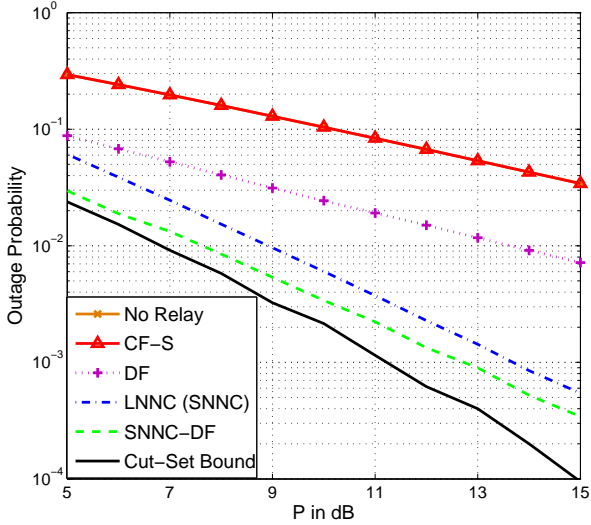}}
\caption{Outage probabilities for a TWRC with Rayleigh fading.}
\label{TWRCOutage}
\end{figure}

Fig.~\ref{TWRCOutage} depicts the outage probabilities with $R_\text{tar1}=2$, $R_\text{tar2}=1$, $P_1=5P$, $P_2=2P$,
$P_3=P$, the geometry of Fig.~\ref{TWRC} and $\alpha=3$.
CF-S has the worst outage probability since it requires that both relay-destination links ($3-1\;\text{and}\;3-2$) are reliable so
that the bin and quantization indices can be recovered at both destinations $1$ and $2$. DF is better than
CF-S, while LNNC (or SNNC) improves on DF. SNNC-DF lets the relay use DF or QF depending on the channel conditions and gains over
about $2$ dB at low power and $1$ dB at high power over LNNC (or SNNC).

\section{Concluding Remarks}
\label{ConRe}
SNNC enables early decoding at nodes, and this enables the use of SNNC-DF. Numerical examples demonstrate that
SNNC-DF shows reasonable gains as compared to DF, CF-S and LNNC in terms of rates and outage probabilities. 


\begin{appendices}

\section{SNNC with joint Decoding}
\label{appA}
After block $B+K\cdot (K-1)$ every node $k\in \set K$ can reliably recover $\mathbf l_B=(l_{1B},\dots,l_{KB})$ via the
multihopping of the last $K(K-1)$ blocks.

Let $\epsilon_1> \epsilon$. Node $k$ tries to find a $(\mathbf {\hat w}^{(k)}_1,\dots ,\mathbf {\hat w}^{(k)}_B )$ and $(\mathbf
{\hat l}^{(k)}_1,\dots,\mathbf {\hat l}^{(k)}_B)$ such that the event (\ref{TypCheck}) occurs for {\em all} $j=1,\dots,B$, where
$\mathbf l_B$ is already known. The difference between {\em joint decoding} and {\em backward decoding} is that
the typicality test is performed jointly over all blocks (see (\ref{eventjoint1})-(\ref{eventjoint3}) below) while it is performed
in only
one block in (\ref{eventblock1})-(\ref{eventblock3}).

{\em Error Probability:} Let $\mathbf 1=(1,\dots,1)$. Assume without loss of generality that $\mathbf w_j=\mathbf 1$ and $\mathbf
l_{j}=\mathbf 1$ for $j=1,\dots,B$. For any $\set S \subset \set K$, define
\begin{align*}
\mathbf w_{(\set S)j}&=[w_{ij}:\; i\in \set S].
\end{align*}
The error events at decoder $k$ are:
\begin{align}
E_{k0}:&\;{\cup^{B}_{j=1}}{\cap_{l_{kj}}}\;E^\text{c}_{0(kj)}(l_{kj}) \label{eventjoint1}\\
E_{k1}:&\; (\cap^B_{j=1}E_{1(kj)}(\mathbf 1,\mathbf 1,\mathbf 1))^\text{c}\label{eventjoint2}\\
E_{k2}:&\; {\cup_{ (\mathbf w^B_{\mathcal {\widetilde D}_k} \ne \mathbf 1, \mathbf w^B_{\cset {\widetilde D}_k}  )}  }
\;{\cup_{\mathbf l^B}}\cap^B_{j=1} \;E_{1(kj)}(\mathbf w_j, \mathbf l_{j-1}, \mathbf l_j )\label{eventjoint3}
\end{align}
The error event $E_k=\cup^2_{i=0} E_{ki}$ at node $k$ thus satisfies
\begin{equation}
\text{Pr}[E_k] \le \text{Pr}[E_{k0}] + \text{Pr}[E_{k1}]+ \text{Pr}[E_{k2}]
\end{equation}
where we have used the union bound.

\begin{table*}[t!]
\begin{center}
\begin{tabular}{r|c c c c c}
  \hline
  Block   &       1            &        2           &       $\cdots$   &            $B$        & $B+1$ \\
\hline
  $X_1$ & $\mathbf x_{11}(w_{1})$ & $\mathbf x_{12}(w_{2})$    &  $\cdots$     &  $\mathbf x_{1B}(w_{B})$ & $\mathbf
x_{1(B+1)}(1)$  \\
  $X_2$ & $\mathbf x_{21}(1)$     & $\mathbf x_{22}(w_1)$ &  $\cdots$   &  $\mathbf x_{2B}(w_{(B-1)})$ & $\mathbf
x_{2(B+1)}(w_{B})$ \\
$X_3$ & $\mathbf x_{31}(1)$   & $\mathbf x_{32}(l_1)$ & $\cdots$   &  $\mathbf x_{3B}(l_{B-1})$ & $\mathbf x_{3(B+1)}(l_{B})$\\ 
$\hat Y_3$  & $\mathbf {\hat y}_{31}(l_{1}|1)$ & $\mathbf {\hat y}_{32}(l_{2}|l_1)$ &$\cdots$   & $\mathbf {\hat
y}_{3B}(l_{B}|l_{B-1})$& $\mathbf {\hat y}_{3{B+1}}(l_{B+1}|l_{B})$\\
\hline
\end{tabular}
\end{center}
\caption{Coding scheme for the two-relay channel without block Markov coding at the source.}
\label{WOBMC}
\end{table*}

$\text{Pr}[E_{k0}]$ can be made small with large $n$ as long as (see (\ref{rdbound}))
\begin{equation}
\label{rdboundjoint}
\hat R_k > I( \hat Y_{k}; Y_{k}|X_k)+\delta_{\epsilon}(n).
\end{equation}
Also, we have 
\begin{align}
\text{Pr}[E_{k1}]&=\text{Pr}[(\cap^B_{j=1}E_{1(kj)}(\mathbf 1,\mathbf 1,\mathbf 1))^\text{c}]\notag\\
&=\text{Pr}[\cup^B_{j=1}E^\text{c}_{1(kj)}(\mathbf 1,\mathbf 1,\mathbf 1)]\notag\\
&{\le} \sum^B_{j=1}\text{Pr}[E^\text{c}_{1(kj)}(\mathbf 1,\mathbf 1,\mathbf 1)]\notag\\
&\overset{(a)}{\le} B\cdot \delta_{\epsilon_1}(n)\notag\\
&=\delta_{\epsilon_1}(n, B) \label{NottypicalUnion}
\end{align}
where $(a)$ follows because $\text{Pr}[E^\text{c}_{1(kj)}(\mathbf 1,\mathbf 1,\mathbf 1)]\le \delta_{\epsilon_1}(n)$, which goes
to zero as $n\rightarrow \infty$, for $j=1,\dots, B $ \cite{Roche01}. 

To bound $\text{Pr}\big[E_{k2} \big]$,  for each $(\mathbf w_j,\mathbf l_{j-1})$, we define 
\begin{align}
\mathcal S_j(\mathbf w_j,\mathbf l_{j-1})=\{i\in \set K: w_{ij}\ne 1\;\text{or}\; l_{i(j-1)}\ne 1\} 
\end{align}
and write $\set S_j=\mathcal S_j(\mathbf w_j,\mathbf l_{j-1})$. Observe that for $j=1,\dots, B:$
\begin{itemize}
 \item $(\mathbf X_{\set S_j}, \mathbf {\hat Y}_{\set S_j})$ is independent of $(\mathbf X_{\cset S_j}, \mathbf {\hat
Y}_{\cset S_j}, \mathbf Y_{kj} )$ in the random coding experiment;
\item the $(X_{ij}, {\hat Y}_{ij}),\; i \in \set S_j,$ are mutually independent.
\end{itemize}
We have (see (\ref{PEISJ}) and (\ref{ISJ})):
\begin{align}
\label{setDistinction}
\text{Pr}\big[E_{1(kj)}(\mathbf w_j, \mathbf l_{j-1}, \mathbf l_j)\big]\le P_{(kj)}(\set S_j)
\end{align}
where
\begin{align}
\label{setDistinction1}
P_{(kj)}(\set S_j)=&\left\{
\begin{array}{ll}
2^{-n(I_{\set S_j}-\delta_{\epsilon_1}(n))} & \text{if} \;\set S_j\ne\emptyset\\
1& \text{otherwise}
\end{array}
\right.
\end{align}
and $\delta_{\epsilon_1}(n)\rightarrow 0$ as $n\rightarrow \infty$.

By the union bound, we have
\begin{align}
&\text{Pr}[E_{k2}]\le\sum_{\left(\mathbf w^B_{\mathcal {\widetilde D}_k} \ne \mathbf 1, \mathbf w^B_{\cset
{\widetilde D}_k}\right) }\sum_{\mathbf l^{B-1}}\text{Pr}[\cap^B_{j=1}E_{1(kj)}\left(\mathbf w_j,
\mathbf l_{j-1},\mathbf l_{j}\right)]\notag\\
&\overset{(a)}{=} \sum_{\left(\mathbf w^B_{\mathcal {\widetilde D}_k} \ne \mathbf 1, \mathbf w^B_{\cset
{\widetilde D}_k}\right) }\sum_{\mathbf l^{B-1}} \prod^B_{j=1}\text{Pr}[E_{1(kj)}\left(\mathbf w_j,
\mathbf l_{j-1},\mathbf l_{j}\right)]\notag\\
&\overset{(b)}{\le} \left[ {\sum_{\mathbf w^B, \mathbf l^{B-1}}}\prod^B_{j=1} \text{Pr}[E_{1(kj)}\left(\mathbf w_j, \mathbf
l_{j-1},\mathbf l_{j}\right)]\right]\notag\\
&\quad\quad\quad\quad\quad\quad-\prod^B_{j=1} \text{Pr}[E_{1(kj)}\left(\mathbf 1, \mathbf 1,\mathbf 1\right)]\notag\\
&\overset{(c)}{\le} \left[ {\sum_{\mathbf w^B, \mathbf l^{B-1}}}\prod^B_{j=1} P_{(kj)}(\set S_j)\right]
-(1-\delta_{\epsilon_1}(n,B))\notag\\
&\overset{(d)}{=}\left[ {\prod^B_{j=1}}\;\;{\sum_{\mathbf w_j, \mathbf
l_{j-1}}}P_{(kj)}(\set S_j)\right]-(1-\delta_{\epsilon_1}(n,B))\notag\\
&\overset{(e)}{<} {\prod^B_{j=1}} \left(1+\underset{\set S\ne \emptyset}{\sum_{\set S: k\in \cset S}}\;\; \underset{\set
S_j(\mathbf w_j,\mathbf l_{j-1})=\set S}{\sum_{ (\mathbf w_j, \mathbf l_{j-1})\ne(\mathbf 1, \mathbf  1) :}} 2^{-n(I_{ \set 
S}-\delta_{\epsilon_1}(n ))}\right)\notag\\
&\quad\quad\quad\quad-(1-\delta_{\epsilon_1}(n,B))\notag\\
&\overset{(f)}{<} \left(1+\underset{\set S\ne \emptyset}{\sum_{\set S: k\in \cset S}} 3^{|\set S|}2^{n\left (R_\set S+\hat R_\set
S) - (I_{\set S}-\delta_{\epsilon_1}(n)) \right) }\right)^B\notag\\
&\quad\quad\quad\quad-(1-\delta_{\epsilon_1}(n,B))
\end{align}
where
\begin{enumerate}[(a)]
\item follows because the codebooks are independent and the channel is memoryless
\item follows by adding $\left(\mathbf w^B_{\mathcal {\widetilde D}_k} = \mathbf 1, \mathbf w^B_{\cset {\widetilde
D}_k}\right)$ to the sum
\item follows from (\ref{setDistinction}) and because (see (\ref{NottypicalUnion}))
\begin{align}
\text{Pr}[\cap^B_{j=1}E_{1(kj)}(\mathbf 1,\mathbf 1,\mathbf 1)]&=\prod^B_{j=1} \text{Pr}[E_{1(kj)}(\mathbf 1,\mathbf 1,\mathbf
1)]\notag\\
&=1-\text{Pr}[(\cap^B_{j=1}E_{1(kj)}(\mathbf 1,\mathbf 1,\mathbf1))^\text{c}]\notag\\
&\ge 1-\delta_{\epsilon_1}(n,B) \label{typicalUnion}
\end{align}
\item follows because $P_{(kj)}(\set S_j)$ depends only on $\mathcal S_j$ which in turn depends only on $(\mathbf
w_j,\mathbf l_{j-1})$
\item follows from (\ref{setDistinction1})
\item follows from (\ref{ErrorBlock4}).
\end{enumerate}

Performing the same steps as in (\ref{Motzkin}) and (\ref{rate2}), we require
\begin{align}
R_{\set S}&<I^\set K_{\set S}(k)
 \label{jointbounds}
\end{align}
for all subsets $\mathcal S \subset \mathcal K $ such that $k\in \cset S$ and $\mathcal S \ne \emptyset$. We can again split the
bounds in (\ref{jointbounds}) into two classes:
\begin{align}
\text{Class}\; 1: &\;\set S \cap \set {\wt D}_k \ne \emptyset\\
\text{Class}\; 2: & \;\set S \cap \set {\wt D}_k = \emptyset \;\text{or equivalently}\; \set S \subseteq \cset {\wt
D}_k \label{constraintj2}
\end{align}
and show that the constraints in (\ref{constraintj2}) at node $k$ are redundant with the same argument used for backward decoding.
By the union bound, the error probability for all destinations tends to zero as $n\rightarrow \infty$ if the rate tuple
$(R_1,\dots,R_K)$ satisfies (\ref{Theo1}) for all subsets $\mathcal S \subset \mathcal K$ such that $k\in \cset S $ and 
$\set S\ne\emptyset,$ and for any joint distribution that factors as (\ref{distribution1}).


\section{Backward Decoding for the Two-Relay Channel without Block Markov Coding}
\label{appB}

The coding scheme is the same as in Example \ref{exp:example4}, except that no BMC is used (see Table~\ref{WOBMC}). We show how 
to recover the rate (\ref{MarkovBack}) with independent inputs and with $2$ different backward decoders.

{\em Decoding at Relays}:
  
$1)$ Node $2$. For $j=1,\dots,B$, node 2 tries to find a $\hat w_{j}$ that satisfies
\begin{align}
(\mathbf x_1\left({\hat w_{j}}), \mathbf x_2({w_{j-1}}), \mathbf y_{2j} \right)\in \set T^{n}_\epsilon\left(P_{X_1X_2Y_2}\right).
\end{align}
Node 2 can reliably decode $w_{j}$ if
\begin{align}
R<I(X_1;Y_2|X_2)-\delta_\epsilon(n) \label{relayBD}
\end{align}
where $\delta_\epsilon(n) \rightarrow 0$ as $n\rightarrow \infty$ (see \cite{Roche01}).

$2)$ Node $3$. For $j=1,\dots, B+1$, node $3$ finds an $l_{j}$ such that 
\begin{align}
\label{rd3}
(\mathbf {\hat y}_{3j}(l_{j}|l_{j-1}),& \mathbf x_{3j}(l_{j-1}), \mathbf y_{3j})\in
\set T^{n}_\epsilon(P_{\hat Y_3X_3Y_3})
\end{align}
if
\begin{align}
\hat R > I(\hat Y_3;Y_3|X_3)+\delta_{\epsilon}(n)
\end{align}
where $\delta_\epsilon(n) \rightarrow 0$ as $n\rightarrow \infty$ (see \cite{Roche01}).

{\em Backward Decoding at the destination:} Let $\epsilon_1>\epsilon$.

\textbf{\em Decoder 1:}

$1)$ Multihop $l_{B+1}$ to node $4$ in blocks $B+2$ to $B+3$. 

$2)$ For $j=B,\dots,1$, node $4$ declares $( w_{j}, l_{j} )=(\hat w_{j}, 
\hat l_{j} )$, if there is a unique pair $(\hat w_{j}, \hat l_{j})$ satisfying the following typicality checks in both blocks 
$j+1$ and $j$:
\begin{align}
&\left(\mathbf x_{1(j+1)}({w_{j+1}}), \mathbf x_{2(j+1)}({\hat w_{j}}), \mathbf x_{3(j+1)}({\hat l_{j}}), \right.
\notag\\&\quad \left.\mathbf {\hat y}_{3(j+1)}(l_{j+1}|\hat l_{j}), \mathbf y_{4(j+1)}\right)\in
\set T^{n}_{\epsilon_1}\left(P_{X_1X_2X_3\hat Y_3Y_4}\right)
\end{align}
and
\begin{align}
(\mathbf x_{1j}({\hat w_{j}}),\mathbf y_{4j} )\in \set T^{n}_{\epsilon_1}\left(P_{X_1Y_4}\right)
\end{align}
where $w_{j+1}$ and $l_{j+1}$ have already been reliably decoded from the previous block $j+1$.

Similar analysis as in Theorem~\ref{thm:theorem1} shows that node 4 can reliably recover $(w_{j}, l_{j})$ if 
\begin{align}
R&<I(X_1;Y_4)+I(X_2;\hat Y_3Y_4|X_1X_3)\label{rateBD1}\\
R&<I(X_1X_2X_3;Y_4)-I(\hat Y_3;Y_3|X_1X_2X_3Y_4)\label{rateBD2}\\
0&\le I(X_3;Y_4|X_1X_2)-I(\hat Y_3;Y_3|X_1X_2X_3Y_4) \label{constraintBD}
\end{align}
If the constraint (\ref{constraintBD}) is violated, then the rate bound~(\ref{rateBD2}) becomes
\begin{align}
R&<I(X_1X_2;Y_4)
\end{align}
which is a stronger bound than (\ref{rateBD1}) and can be achieved with SNNC-DF by treating $X_3$ as
noise. Thus, we may ignore (\ref{constraintBD}).

\textbf{\em Decoder 2:}

$1)$ Multihop $l_{B+1}$ and $l_{B}$ to node $4$ in blocks $B+2$ to $B+5$.

$2)$ For $j=B,\dots,1$, node 4 declares $( w_{j}, l_{j-1} )=(\hat w_{j}, \hat l_{j-1} )$, if
there is a unique pair $(\hat w_{j}, \hat l_{j-1})$ satisfying the following typicality checks in both blocks $j+1$ and $j$:
\begin{align}
(\mathbf x_{1(j+1)}&({w_{j+1}}), \mathbf x_{2(j+1)}({\hat w_{j}}), \mathbf x_{3(j+1)}({l_{j}}),\\& \mathbf {\hat 
y}_{3(j)}(l_{j+1}| l_{j}), \mathbf y_{4(j+1)} )\in \set T^{n}_{\epsilon_1}\left(P_{X_1X_2X_3\hat Y_3Y_4}\right)
\end{align}
and
\begin{align}
&\left(\mathbf x_{1j}({\hat w_{j}}),  \mathbf x_{3j}({\hat l_{j-1}}), \mathbf {\hat y}_{3j}(l_{j}|\hat l_{j-1}), \mathbf 
y_{4j}\right)\in \set T^{n}_{\epsilon_1}\left(P_{X_1X_3\hat Y_3Y_4}\right)
\end{align}
where $w_{j+1}$, $l_{j}$ and $l_{j+1}$ have already been reliably decoded from the previous block $j+1$.

Node $4$ can reliably recover $(w_{j}, l_{j-1})$ if 
\begin{align}
R&<I(X_1X_2;\hat Y_3Y_4|X_3)\label{de21}\\
R&<I(X_1X_2X_3;Y_4)-I(\hat Y_3;Y_3|X_1X_2X_3Y_4)\label{de22}\\
0&\le I(X_3;Y_4|X_1)-I(\hat Y_3;Y_3|X_1X_3Y_4) \label{de23}
\end{align}
If the constraint (\ref{de23}) is violated, then the rate bound~(\ref{de22}) becomes
\begin{align}
R&<I(X_1;Y_4)+I(X_2;\hat Y_3Y_4|X_1X_3)
\end{align}
and the resulting $R$ can be achieved by using decoder $1$ (see (\ref{rateBD1})). Thus, with the combination of both 
decoders, we may ignore (\ref{de23}) and achieve the rate (\ref{MarkovBack}).

\begin{remark}
Sliding window decoding with 2 different decoders also recovers the rate (\ref{MarkovBack}) for independent $X_1$ and $X_2$ and 
enjoys a smaller decoding delay.
\end{remark}

\section{Rates and Outage for Gaussian Networks}

\label{appC}
In the following, let $C(x)=\log_2(1+x)$, $x\ge 0$. 

\subsection{Relay Channels}

\subsubsection{No Fading}
The achievable rates $R$ with DF and CF-S are given in \cite{Kramer04}. The SNNC and LNNC rates are simply the CF-S rate. The
SNNC-DF rate is the larger of the SNNC and DF rates.

\subsubsection{Slow Rayleigh Fading}
Define the events
\begin{align}
D_\text{DF}&=\left\{R_\text{tar}<C\left(|G_{12}|^2P_1(1-|\beta|^2)\right)\right\}\notag\\
D_{\text{CF-S}1}&=\left\{R_{2(\text{bin})}<C\left(\frac{ {|G_{23}|^2P_2}}{1+ |G_{13}|^2P_1}\right)\right\}\notag\\
D_{\text{CF-S}2}&=\left\{R_{2(\text{bin})}  \ge C\left(\frac{1}{\hat \sigma^2_2} +\frac{ {|G_{12}|^2P_1} }{\hat
\sigma^2_2(1+|G_{13}|^2P_1)} \right) \right\}\notag\\
D_{\text{SNNC}}&=\left\{ \hat \sigma^2_2 \ge \frac{1}{ |G_{23}|^2P_2}\right\}
\end{align}
where $|\beta|^2$ is the fraction of power allocated by source $1$ to sending new messages. The optimal $\beta$,
$R_{2(\text{bin})}$ and $\hat \sigma^2_2$ are calculated numerically.

The DF, CF-S, SNNC and SNNC-DF rates are
\begin{align}
&R_\text{DF}=a_1\notag\\
&R_\text{CF-S}=b_1\notag\\
&R_\text{SNNC}=c_1\notag\\
&R_\text{SNNC-DF}=\left\{\begin{array}{ll}
R_\text{DF} &  \text{if} \; D_\text{DF}\; \text{occurs}\\
R_\text{SNNC} & \text{otherwise}
              \end{array}\right.
\end{align}
where 
\begin{align}
a_1&=\text{min}\left\{C\left( |G_{12}|^2P_1(1-|\beta|^2) \right), \notag\right.\\
&\left.\quad C\left( |G_{13}|^2P_1
+ |G_{23}|^2P_2+2\Re\{\beta G_{13}G^\ast_{23}\}\sqrt{P_1P_2}\right)\label{rho12}\right\}\notag\\
\notag b_1&=\left\{
\begin{array}{ll}
C\left(\frac{|G_{12}|^2P_1}{1+\hat \sigma^2_2 }+{|G_{13}|^2P_1}\right) &
\text{if} \; D_{\text{CF-S}1} \cap D_{\text{CF-S}2}\\
C\left(|G_{13}|^2P_1\right) & \text{if}\; D_{\text{CF-S}1} \cap D^\text{c}_{\text{CF-S}2}\\
C\left(\frac{|G_{13}|^2P_1} {1+|G_{23}|^2P_2}  \right) & \text{otherwise}
\end{array}\right.\\
c_1&=\left\{
\begin{array}{ll}
\text{min}\; \left\{C\left(|G_{13}|^2 P_1+|G_{23}|^2 P_2\right)-C(\frac{1}{\hat \sigma^2_2}), \right.\\
\left.C\left(\frac{|G_{12}|^2P_1}{1+\hat \sigma^2_2}+|G_{13}|^2P_1\right)\right\} \quad\; \text{if}\; D_{\text{SNNC}}\\
C\left(\frac{|G_{13}|^2P_1}{1+|G_{23}|^2P_2}  \right) \quad\quad\quad\quad\quad\quad\text{otherwise} 
\end{array}
\right.
\end{align}
and $\Re\{x\}$ is the real part of $x$ and $x^\ast$ is the complex conjugate of $x$.

\begin{remark}
\label{rem:remark7}
For SNNC, event $D_\text{SNNC}$ means that 
\begin{align}
 I(X_2;Y_3|X_1)-I(\hat Y_2;Y_2|X_1X_2Y_3)\ge 0
\end{align}
and  the destination can reliably recover $X_2$ and $\hat Y_2$ {\em jointly} which helps to decode $X_1$. Otherwise the
destination should treat $X_2$ as noise to get a better rate (see Theorem~\ref{thm:theorem1}). Similarly, for CF-S the
events $D_{\text{CF-S}1}$ and $D_{\text{CF-S}2}$ mean that both $X_2$ and $\hat Y_2$ can be decoded in a {\em step-by-step}
fashion \cite{Cover01}. If $D_{\text{CF-S}1}$ and $D^\text{c}_{\text{CF-S}2}$ occur, then $X_2$ can be recovered which removes
interference at the receiver. Otherwise the relay signal should be treated as noise.

As recognized in \cite{Laneman01}, one drawback of DF is that if the source-relay link happens to be weak and
the relay tries to decode, then the rate suffers. Hence the relay should decode only if the source-relay link is strong
enough to support $R_\text{tar}$, i.e., if event $D_\text{DF}$ occurs. Otherwise, the relay should perform CF-S or QF. Different
choices of relay operations depending on the channel conditions lead to the achievable rates with SNNC-DF.
\end{remark}

The outage probabilities are as follows:
\begin{align}
&P^\text{out}_\text{DF}=\text{Pr}[R_\text{DF}<R_\text{tar}]\notag\\
&P^\text{out}_\text{CF-S}=\text{Pr}[R_\text{CF-S}<R_\text{tar}]\notag\\
&P^\text{out}_\text{SNNC}=\text{Pr}[R_\text{SNNC}<R_\text{tar}]\notag\\ 
&P^\text{out}_\text{SNNC-DF}=\text{Pr}[R_\text{SNNC-DF}<R_\text{tar}]\label{OUTRELAY}
\end{align}

\subsection{Two-Relay Channels}
\subsubsection{No Fading}
The achievable DF rates are \cite[Theorem 1]{Kramer04}
\begin{align}
R_\text{DF}<\text{max}\;\{R_{\text{DF}1},R_{\text{DF}2} \}
\end{align}
where
\begin{align}
&R_{\text{DF}1}=\text{min } \left\{a_{21}, a_{22}, a_{23}\right\}\notag\\
&R_{\text{DF}2}=\text{min } \left\{b_{21}, b_{22}, b_{23} \right\}\label{a211}
\end{align}
with
\begin{align}
&a_{21}= C\left(|\beta_1|^2|G_{12}|^2 P_1 \right)\notag\\
&a_{22}=C\left((1-|\beta_3|^2)|G_{13}|^2P_1+|\gamma_1|^2|G_{23}|^2P_2 \right.\notag\\
&\left.\quad\quad\quad+2\Re\{\beta_2G_{13}(\gamma_1 G_{23})^\ast \}\sqrt{P_1P_2} \right)\notag\\
&a_{23}=C\left( |G_{14}|^2 P_1 + |G_{24}|^2 P_2 + |G_{34}|^2 P_3\right.\notag\\
&\;+\left( 2\Re\{ \beta_2G_{14}(\gamma_1G_{24})^\ast \} +2\Re\{ \beta_3G_{14}(\gamma_2G_{24})^\ast \}\right) \sqrt{P_1P_2}\notag\\
 &\left.\quad+ 2\Re\{\beta_3G_{14}G^\ast_{34} \} \sqrt{P_1P_3} + 2\Re\{\gamma_2G_{24}G^\ast_{34}\} \sqrt{P_2P_3}\right)\notag\\
&b_{21}=C\left(|\beta_1|^2|G_{13}|^2 P_1 \right)\notag\\
&b_{22}=C\left((1-|\beta_3|^2)|G_{12}|^2P_1+|\gamma_1|^2|G_{32}|^2P_3 \right.\notag\\
&\left.\quad\quad\quad+2\Re\{\beta_2G_{12}(\gamma_1 G_{32})^\ast \}\sqrt{P_1P_3} \right)\notag\\
&b_{23}=C\left( |G_{14}|^2 P_1 + |G_{24}|^2 P_2 + |G_{34}|^2 P_3\right.\notag\\
&\;+\left( 2\Re\{ \beta_2G_{14}(\gamma_1G_{34})^\ast \} +2\Re\{ \beta_3G_{14}(\gamma_2G_{34})^\ast \}\right) \sqrt{P_1P_3}\notag\\
 &\left.\quad+ 2\Re\{\beta_2G_{14}G^\ast_{24} \} \sqrt{P_1P_2} + 2\Re\{\gamma_1G_{24}G^\ast_{34}\} \sqrt{P_2P_3}\right)\label{a21}
\end{align}
where $\sum^3_{i=1}|\beta_i|^2=1$ and $\sum^2_{i=1}|\gamma_i|^2=1$ and the optimal power allocation parameters are calculated
numerically.

The CF-S rates are (see \cite[Theorem 2]{Kramer04} with $U_i=0$, $i=2,3$)
\begin{align}
R_\text{CF-S}<c_{21}
\end{align}
subject to
\begin{align*}
g_2\le d_2,\;h_2\le e_2,\;i_2\le f_2
\end{align*}
where
\begin{align}
c_{21}&=C\left(\frac{|G_{12}|^2P_1}{1+\hat \sigma^2_2}+\frac{|G_{13}|^2P_1}{1+\hat \sigma^2_3}+{|G_{14}|^2P_1}\right)\notag\\
d_2 &=C\left(\frac{|G_{24}|^2P_2}{1+|G_{14}|^2P_1}  \right)\notag\\
e_2&=C\left(\frac{ |G_{34}|^2P_3}{1+|G_{14}|^2P_1} \right)\notag\\
f_2&=C\left(\frac{ |G_{24}|^2P_2+|G_{34}|^2P_3}{1+ |G_{14}|^2P_1} \right)\notag\\
g_2&=C\left( \frac{1}{\hat \sigma^2_2}+\frac{ |G_{12}|^2P_1}{\hat \sigma^2_2(1+\frac{ |G_{13}|^2P_{1}}{ (1+\hat
\sigma^2_3)}+|G_{14}|^2P_{1} )} \right)\notag\\
h_2&=C\left(\frac{1}{\hat  \sigma^2_3}+\frac{ |G_{13}|^2{P_1} } {\hat \sigma^2_3
(1+\frac{ |G_{12}|^2 P_1}{1+\hat \sigma^2_2}+|G_{14}|^2P_1 ) } \right)\notag\\
i_2&=C\left( \frac{1+\hat \sigma^2_2+\hat \sigma^2_3}{\hat \sigma^2_2\hat \sigma^2_3}\right.\notag\\
&\left.\quad\quad+\frac{|G_{12}|^2P_1(1+\hat \sigma^2_3)+|G_{13}|^2P_1(1+\hat \sigma^2_2)}{\hat \sigma^2_2\hat
\sigma^2_3(1+|G_{14}|^2P_1)} \right).\label{c21}
\end{align}
The optimal $\hat \sigma^2_2$ and $\hat \sigma^2_3$ are calculated numerically.

Referring to Theorem~\ref{thm:theorem1}, the achievable SNNC rates are
\begin{align}
R_\text{SNNC}<\text{min}\; \left\{c_{21}, j_{21}, j_{22}, j_{23}\right\}
\end{align}
where
\begin{align}
j_{21}&=C\left(|G_{14}|^2P_1+|G_{24}|^2P_2+\frac{|G_{13}|^2P_1+|G_{23}|^2P_2}{1+\hat\sigma^2_3}\right.\notag\\
&\left.\quad\quad\quad+\frac{P_1P_2(|G_{13}|^2|G_{24}|^2+|G_{14}|^2|G_{23}|^2)}{1+\hat\sigma^2_3}\right. \notag\\ 
&\left.\quad\quad\quad-\frac{2\Re\{ G_{13}G_{24}G^\ast_{14}G^\ast_{23}\}P_1P_2}{1+\hat\sigma^2_3}   \right)
-C\left(\frac{1}{\hat\sigma^2_2}\right)\notag\\
j_{22}&=C\left( |G_{14}|^2P_1+|G_{34}|^2P_3+\frac{|G_{12}|^2P_1+|G_{32}|^2P_3}{1+\hat\sigma^2_3}\right.\notag\\
&\left.\quad\quad\quad+\frac{P_1P_3(|G_{12}|^2|G_{34}|^2+|G_{14}|^2|G_{32}|^2)}{(1+\hat\sigma^2_2)}\right.\notag\\
&\left.\quad\quad\quad -\frac{2\Re\left\{ G_{12}G_{34}G^\ast_{14}G^\ast_{32}\right\} P_1P_3}{(1+\hat\sigma^2_2) } 
\right)-C\left(\frac{1}{ \hat\sigma^2_3}\right)\notag\\
j_{23}&=C\left(|G_{14}|^2P_1+|G_{24}|^2P_2+|G_{34}|^2P_3\right)\notag\\
&\quad -C\left(\frac{1+\hat \sigma^2_2+\hat \sigma^2_3}{\hat \sigma^2_2\hat\sigma^2_3}\right).
\end{align}
where $c_{21}$ is defined in (\ref{c21}). The optimal $\hat \sigma^2_2$ and $\hat \sigma^2_3$ are calculated numerically.

If one relay uses DF and the other uses QF, rates satisfying
\begin{align}
R_\text{DQF} < \text{max}\;\left\{ R_{\text{DQF}1},R_{\text{DQF}2} \right\}
\end{align}
can be achieved, where
\begin{align*}
&R_{\text{DQF}1}=\text{min}\;\left\{k_{21}, k_{22}, k_{23}\right\}\\
&R_{\text{DQF}2}=\text{min}\;\left\{l_{21}, l_{22}, l_{23} \right\}
\end{align*}
with
\begin{align}
k_{21}&=C\left( \frac{|G_{12}|^2P_1(1-|\theta|^2)}{1+|G_{32}|^2P_3} \right)\notag\\
k_{22}&=C\left(|G_{14}|^2P_1+|G_{24}|^2P_2+ 2\Re\{ \theta G_{14}G^\ast_{24}\}\sqrt{P_1P_2}\right.\notag\\
&+\frac{|G_{13}|^2P_1+|G_{23}|^2P_2+2\Re\{ \theta G_{13}G^\ast_{23}\}\sqrt{P_1P_2}}{1+\hat\sigma^2_3}\notag \\
&+ \frac{(1-|\theta|^2)P_1P_2( |G_{13}|^2|G_{24}|^2+ |G_{14}|^2|G_{23}|^2)}{{1+\hat\sigma^2_3}}\notag\\
&\left.- \frac{(1-|\theta|^2)P_1P_2 \cdot 2\Re\{G_{13}G_{24}G^\ast_{14}G^\ast_{23}\} )}{{1+\hat\sigma^2_3}}\right)\notag\\
k_{23}&=C\left(|G_{14}|^2P_1+|G_{24}|^2P_2 +|G_{34}|^2P_3\right.\notag\\
&\left.\quad\quad+2\Re\{ \theta G_{14}G^\ast_{24}\}\sqrt{P_1P_2}\right)-C\left(\frac{1}{\hat \sigma^2_3}\right)\label{k21}
\end{align}
and 
\begin{align}
l_{21}&=C\left( \frac{|G_{13}|^2P_1(1-|\theta|^2)}{1+|G_{23}|^2P_2} \right)\notag\\
l_{22}&=C\left(|G_{14}|^2P_1+|G_{34}|^2P_3+ 2\Re\{ \theta G_{14}G^\ast_{34}\}\sqrt{P_1P_3}\right.\notag\\
&+\frac{|G_{12}|^2P_1+|G_{32}|^2P_3+2\Re\{ \theta G_{12}G^\ast_{32}\}\sqrt{P_1P_3}}{1+\hat\sigma^2_2}\notag \\
&+ \frac{(1-|\theta|^2)P_1P_3( |G_{12}|^2|G_{34}|^2+ |G_{14}|^2|G_{32}|^2)}{{1+\hat\sigma^2_2}}\notag\\
&\left.- \frac{(1-|\theta|^2)P_1P_3 \cdot 2\Re\{G_{12}G_{34}G^\ast_{14}G^\ast_{32}\} )}{{1+\hat\sigma^2_2}}\right)\notag\\
l_{23}&=C\left(|G_{14}|^2P_1+|G_{24}|^2P_2 +|G_{34}|^2P_3\right.\notag\\
&\left.\quad\quad+2\Re\{ \theta G_{14}G^\ast_{34}\}\sqrt{P_1P_3}\right)-C\left(\frac{1}{\hat \sigma^2_2}\right)\label{l21}
\end{align}
where $0\le |\theta|^2 \le 1$ and the optimal $\theta$, $\hat \sigma^2_2$ and $\hat \sigma^2_3$ for $R_{\text{DQF}1}$ and
$R_{\text{DQF}2}$ are calculated numerically.

Referring to Theorem~\ref{thm:theorem2}, SNNC-DF achieves rates satisfying
\begin{align}
R_\text{SNNC-DF}<\text{max}\:\{R_\text{DF}, R_\text{DQF}, R_\text{SNNC}\}.
\end{align}

\subsubsection{Slow Rayleigh Fading}
Define the events
\begin{align}
\notag D_{\text{DFV} }&=\left\{
\begin{array}{ll}
R_\text{tar}<V_{21}\\
R_\text{tar}<V_{22}
\end{array}
\right\}\\
\notag  D_{\text{DF}1}&=\left\{ \:\; R_\text{tar}<k_{21}\right\}\\
\notag  D_{\text{DF}2}&=\left\{ \:\;R_\text{tar}<l_{21}\right\}\\
\notag  D_{\text{CF-S}1}&=\left\{
\begin{array}{ll}
R_{2(\text{bin})}<d_2 \\
R_{3(\text{bin})}<e_2\\
R_{2(\text{bin})} + R_{3(\text{bin})}<f_2\\
\end{array}
\right\}\\
\notag  D_{\text{CF-S}2}&=\left\{
\begin{array}{ll}
 R_{2(\text{bin})} \ge g_2 \\
 R_{3(\text{bin})} \ge h_2 \\
 R_{2(\text{bin})} + R_{3(\text{bin})} \ge i_2
\end{array}
\right\}\\
\notag  D_{\text{SNNC}1}&=\left\{
\begin{array}{ll}
|G_{24}|^2P_2   + \frac{ |G_{23}|^2P_2}{1+\hat \sigma^2_3} \ge\frac{1}{\hat\sigma^2_2}\\
|G_{34}|^2P_3  + \frac{  |G_{32}|^2P_3}{1+\hat \sigma^2_2}\ge \frac{1}{\hat\sigma^2_3}\\
|G_{24}|^2P_2 + |G_{34}|^2P_3\ge \frac{1}{\hat\sigma^2_2}+\frac{1}{\hat\sigma^2_3}+\frac{1}{\hat\sigma^2_2\hat\sigma^2_3}
\end{array}
\right\}\\
\notag D_{\text{SNNC}2}&=\left\{\hat\sigma^2_2 \ge \frac{1+|G_{32}|^2P_3+|G_{34}|^2P_3}{ |G_{24}|^2P_2}\right\}\\
D_{\text{SNNC}3}&=\left\{\hat\sigma^2_3 \ge \frac{1+|G_{23}|^2P_2+|G_{24}|^2P_2}{ |G_{34}|^2P_3}\right\}
\end{align}
where $\{V_{21}, V_{22}, V_{23}\} $ takes on the value $\{a_{21}, a_{22}, a_{23}\}$ or $\{b_{21}, b_{22}, b_{23}\}$ (see
(\ref{a21}))
and the choice depends on the statistics of the fading coefficients such that the DF outage probability is minimized. 

The DF rates are 
\begin{align}
R_{\text{DF}}=\text{min }& \left\{V_{21}, V_{22}, V_{23}\right\}.
\end{align}
The CF-S rates are
\begin{align}
R_\text{CF-S}=\left\{
\begin{array}{ll}
c_{21}& \text{if} \; D_{\text{CF-S}1} \cap D_{\text{CF-S}2}\\
c_{22}& \text{if} \; D_{\text{CF-S}1} \cap D^\text{c}_{\text{CF-S}2}\\
c_{23}& \text{otherwise}
\end{array}
\right.
\end{align}
where $c_{21}$ is defined in (\ref{c21}) and
\begin{align}
c_{22}&=C\left(|G_{14}|^2P_1\right)\notag\\
c_{23}&=C\left( \frac{ |G_{14}|^2P_1}{1+|G_{24}|^2P_2+|G_{34}|^2P_3} \right).\label{c23}
\end{align}
Observe that if both $D_{\text{CF-S}1}$ and $D_{\text{CF-S}2}$ occur, then both the bin and quantization
indices can be decoded. If only $D_{\text{CF-S}1}$ occurs, then only the bin index can be recovered.

Referring to Theorem~\ref{thm:theorem1} the SNNC rates are
\begin{align}
R_\text{SNNC}=\left\{
\begin{array}{ll}
\text{min}\; \left\{c_{21}, j_{21}, j_{22}, j_{23}\right\} & \text{if} \; D_{\text{SNNC}1}\\
\text{min}\; \left\{m_{21}, m_{22}\right\} & \text{if} \; D^\text{c}_{\text{SNNC}1} \cap D_{\text{SNNC}2}\\
\text{min}\; \left\{q_{21}, q_{22}\right\} & \text{if} \; D^\text{c}_{\text{SNNC}1} \cap D_{\text{SNNC}3}\\
c_{23} & \text{otherwise} 
\end{array}
\right.
\end{align}
where
\begin{align}
m_{21}&=C\left(\frac{P_1(|G_{12}|^2+(1+\hat\sigma^2_2)|G_{14}|^2)+P_1P_3|G_{14}|^2|G_{32}|^2}{|G_{32}
|^2P_3+(1+\hat\sigma^2_2)\left(1+|G_{34}|^2P_3\right)} \right.\notag\\
&\left.\quad\quad\quad +\frac{ P_1P_3(|G_{12}|^2|G_{34}|^2-2\Re\{ G_{12}G_{34}G^\ast_{14}G^\ast_{32}
\})}{|G_{32}|^2P_3+(1+\hat\sigma^2_2)\left(1+ |G_{34}|^2P_3\right)}\right) \notag\\
m_{22}&=C\left(\frac{  |G_{14}|^2P_1 + |G_{24}|^2P_2}{1+ |G_{34}|^2P_3}\right)\notag\\
&\quad\quad\quad-C\left(\frac{1}{\hat\sigma^2_2}+\frac{  |G_{32}|^2P_3}{ \hat
\sigma^2_2\left(1+|G_{34}|^2P_3\right)}\right)\notag\\
q_{21}&=C\left(\frac{P_1(|G_{13}|^2+(1+\hat\sigma^2_3) |G_{14}|^2)+
P_1P_2|G_{14}|^2|G_{23}|^2}{|G_{23}|^2P_2+(1+\hat\sigma^2_3)\left(1+|G_{24}|^2P_2\right) }
\right.\notag\\
&\left.\quad\quad\quad+\frac{P_1P_2(|G_{13}|^2|G_{24}|^2-2\Re\{ G_{13}G_{24}G^\ast_{14}G^\ast_{23}
\})}{|G_{23}|^2P_2+(1+\hat\sigma^2_3)\left(1+ |G_{24}|^2P_2\right)}\right) \notag\\
q_{22}&=C\left(\frac{  |G_{14}|^2P_1 + |G_{34}|^2P_3}{1+ |G_{24}|^2P_2}\right)\notag\\
&\quad\quad-C\left(\frac{1}{\hat\sigma^2_3}+\frac{  |G_{23}|^2P_2}{\hat \sigma^2_3\left(1+|G_{24}|^2P_2\right)} \right).
\end{align}
The event $D_{\text{SNNC}1}$ means that both quantization indices can be recovered. The events $D_{\text{SNNC}2}$ and
$D_{\text{SNNC}3}$ mean that only one of the two quantization indices can be decoded.

The SNNC-DF rates are
\begin{align}
R_\text{SNNC-DF}=\left\{
\begin{array}{ll}
R_{\text{DF}} & \text{if}\; D_{\text{DFV}}\\
R_{\text{DQF}1} & \text{if}\; D^\text{c}_{\text{DFV}} \cap D_{\text{DF}1} \\
R_{\text{DQF}2} & \text{if}\; D^\text{c}_{\text{DFV}} \cap D_{\text{DF}2} \\
R_\text{SNNC} & \text{otherwise}
\end{array}
\right.
\end{align}
where (see (\ref{k21}) and (\ref{l21}))
\begin{align*}
&R_\text{DQF1}=\text{min}\;\left\{k_{21}, k_{22}, k_{23}\right\}\\
&R_\text{DQF2}=\text{min}\;\left\{l_{21}, l_{22}, l_{23} \right\}.
\end{align*}
The outage probabilities are as in (\ref{OUTRELAY}).

\subsection{Multiple Access Relay Channels}
\subsubsection{No Fading}
The DF rate region of the Gaussian MARC is the union of all pairs $(R_1,R_2)$ satisfying \cite[Sec. 3]{Sankar02}
\begin{align}
R_1&<R_{\text{DF}1}=\text{min}\;\left\{a_{31}, a_{32} \right\}\notag\\
R_2&<R_{\text{DF}2}=\text{min}\;\left\{b_{31}, b_{32}\right\}\notag\\
R_1+R_2&<R_{\text{DF}3}=\text{min}\;\left\{ c_{31}, c_{32}\right\}\label{RDFMARC}
\end{align}
where
\begin{align}
a_{31}&=C\left( |G_{13}|^2P_1(1-|\beta|^2)\right)\notag\\
a_{32}&=C\left(|G_{14}|^2P_1+|G_{34}|^2 P_3 \right. \notag\\
&\left.\quad + 2\Re\{\beta G_{14} (\theta_1 G_{34})^\ast \}\sqrt{P_1P_3}\right)\notag\\
b_{31}&=C\left(|G_{23}|^2P_2(1-|\gamma|^2)\right)\notag\\
b_{32}&=C\left(|G_{24}|^2P_2+|G_{34}|^2 P_3\right.\notag \\
&\left.\quad + 2\Re\{\gamma G_{24} (\theta_2 G_{34})^\ast \}\sqrt{P_2P_3} \right)\notag\\
c_{31}&=C\left(|G_{13}|^2P_1(1-|\beta|^2)+|G_{23}|^2P_2(1-|\gamma|^2)\right)\notag\\
c_{32}&=C\left( |G_{14}|^2P_1+|G_{24}|^2P_2 + |G_{34}|^2P_3 \right.\notag \notag\\
&\quad\quad +2\Re\{\beta G_{14} (\theta_1 G_{34})^\ast \}\sqrt{P_1P_3}\notag\\ 
&\left.\quad\quad+ 2\Re\{\gamma G_{24} (\theta_2 G_{34})^\ast \}\sqrt{P_2P_3} \right)\label{a31}
\end{align}
where $0\le |\beta|^2, |\gamma|^2\le 1$ and $\sum^2_{i=1}|\theta_i|^2=1$. The optimal power allocation parameters
are calculated numerically.

The achievable CF-S rate region is the union of all pairs $(R_1,R_2)$ satisfying \cite[Sec. 3]{Sankar02}
\begin{align}
R_1&<d_{31} \notag\\
R_2&<e_{31}\notag \\
R_1+R_2&<f_{31}
\end{align}
where
\begin{align}
& d_{31}=C\left(\frac{|G_{13}|^2P_1}{1+\hat \sigma^2_3} +|G_{14}|^2P_1\right) \notag\\
& e_{31}=C\left(\frac{|G_{23}|^2P_2}{1+\hat \sigma^2_3} +|G_{24}|^2P_2\right) \notag\\
& f_{31}=C \left(|G_{14}|^2P_1+|G_{24}|^2P_2+\frac{ |G_{13}|^2P_1+|G_{23}|^2P_2}{1+\hat \sigma^2_3} \right. \notag\\
&\left.+\frac{P_1P_2(|G_{13}|^2|G_{24}|^2+|G_{14}|^2|G_{23}|^2-2\Re\{G_{13}G_{24}G^\ast_{14}G^\ast_{23}\} )}{1+\hat\sigma^2_3}
\right)\label{d31}
\end{align}
for some
\begin{align*}
 &\hat \sigma^2_3\ge  \frac{1+(|G_{13}|^2+|G_{14}|^2)P_1+(|G_{23}|^2+|G_{24}|^2)P_2} {|G_{34}|^2P_3}\\
&+\frac{P_1P_2(|G_{13}|^2|G_{24}|^2+|G_{14}|^2|G_{23}|^2-2\Re\{G_{13}G_{24}G^\ast_{14}G^\ast_{23}\})}{|G_{34}|^2P_3}.
\end{align*}

Referring to Theorem~\ref{thm:theorem1}, the SNNC rate region is the union of all pairs $(R_1,R_2)$ satisfying
\begin{align}
R_1&<\text{min}\;\left\{ d_{31}, g_{31}\right\}\notag \\
R_2&<\text{min}\;\left\{ e_{31}, h_{31}\right\}\notag \\
R_1+R_2&<\text{min}\;\left\{f_{31}, i_{31}\right\}
\end{align}
where $d_{31}, e_{31}$ and $f_{31}$ are defined in (\ref{d31}) and
\begin{align*}
g_{31}&=C\left(|G_{14}|^2P_1+ |G_{34}|^2P_3 \right)-C\left(\frac{1}{\hat \sigma^2_3}\right)\notag\\
h_{31}&=C\left(|G_{24}|^2P_2+ |G_{34}|^2P_3 \right)-C\left(\frac{1}{\hat \sigma^2_3}\right)\notag\\
i_{31}&=C\left( |G_{14}|^2P_1 + |G_{24}|^2P_2 + |G_{34}|^2P_3 \right)-C\left(\frac{1}{\hat \sigma^2_3}\right)
\end{align*}
for some $\hat \sigma^2_3 > \frac{1}{ |G_{34}|^2P_3}$. The SNNC-DF rate region is the union of the SNNC and DF rate regions.

\subsubsection{Slow Rayleigh Fading}
Define the events
\begin{align}
\notag  &D_\text{DF}=\left\{
\begin{array}{ll}
R_{\text{tar}1}<a_{31}\\
R_{\text{tar}2}<b_{31}\\
R_{\text{tar}1}+R_{\text{tar}2}<c_{31}
\end{array}
\right\}\\
\notag  &D_{\text{CF-S}1}=\left\{R_{3(\text{bin})}<C\left(\frac{|G_{34}|^2P_3}{1+|G_{14}|^2P_1+|G_{24}|^2P_2}\right)\right\} \\
\notag  &D_{\text{CF-S}2}=\left\{R_{3(\text{bin})} \ge C\left(\frac{1}{\hat \sigma^2_3} + \frac{ |G_{13}|^2 P_1+|G_{23}|^2P_2
}{\hat \sigma^2_3\left(1+|G_{14}|^2 P_1 + |G_{24}|^2 P_2 \right)} \right.\right.\notag \\
&\quad\quad\quad\quad\quad+\frac{P_1P_2(|G_{13}|^2|G_{24}|^2+|G_{14}|^2|G_{23}|^2)}{  \hat
\sigma^2_3\left(1+|G_{14}|^2P_1+|G_{24}|^2 P_2 \right)}\notag \\
&\left.\left.\quad\quad\quad\quad\quad-\frac{P_1P_2\cdot 2\Re\{G_{13}G_{24}G^\ast_{14}G^\ast_{23}\}}{  \hat
\sigma^2_3\left(1+|G_{14}|^2P_1+|G_{24}|^2 P_2 \right)}\right)\right\}\notag \\
&D_\text{SNNC}=\left\{\hat \sigma^2_3\ge \frac{1}{ |G_{34}|^2P_3}\right\}.
\end{align}

The DF rate region of the Gaussian MARC is the union of all rate
pairs $(R_1,R_2)$ satisfying (\ref{RDFMARC}). The CF-S rate region is the union of all $(R_1,R_2)$ satisfying \cite{Sankar02}
\begin{align}
R_1&<R_{\text{CF}1}=\left\{
\begin{array}{ll}
d_{31} & \text{if}\;D_{\text{CF-S}1} \cap D_{\text{CF-S}2}\\
d_{32} & \text{if} \;D_{\text{CF-S}1} \cap D^\text{c}_{\text{CF-S}2}\\
d_{33} & \text{otherwise}
\end{array}
\right.\\
R_2&<R_{\text{CF}2}=\left\{
\begin{array}{ll}
e_{31}  & \text{if}\;D_{\text{CF-S}1} \cap D_{\text{CF-S}2}\\
e_{32}  & \text{if} \;D_{\text{CF-S}1} \cap D^\text{c}_{\text{CF-S}2}\\
e_{33}  & \text{otherwise}
\end{array}
\right.\\
R_1+R_2&<R_{\text{CF}3}=\left\{
\begin{array}{ll}
f_{31} &\text{if}\;D_{\text{CF-S}_1} \cap D_{\text{CF-S}2}\\
f_{32} &\text{if} \;D_{\text{CF-S}_1} \cap D^\text{c}_{\text{CF-S}2}\\
f_{33} & \text{otherwise}
\end{array}
\right.
\end{align}
where
\begin{align}
d_{32}&=C\left(|G_{14}|^2P_1\right)\notag\\
d_{33}&=C\left(\frac{|G_{14}|^2P_1}{1+|G_{34}|^2P_3}\right)\notag\\
e_{32}&=C\left(|G_{24}|^2P_2\right)\notag\\
e_{33}&=C\left(\frac{|G_{24}|^2P_2}{1+|G_{34}|^2P_3}\right)\notag\\
f_{32}&=C\left( |G_{14}|^2P_1+|G_{24}|^2P_2\right)\notag\\
f_{33}&=C\left(\frac{ |G_{14}|^2P_1+ |G_{24}|^2P_2}{1+|G_{34}|^2P_3}\right).
\end{align}
If both $D_{\text{CF-S}1}$ and $D_{\text{CF-S}2}$ occur, then the relay bin and quantization indices can
be decoded. If only $D_{\text{CF-S}1}$ occurs, then only the bin index can be recovered.

Referring to Theorem~\ref{thm:theorem1}, the SNNC rate region is the union of all pairs $(R_1,R_2)$ satisfying
\begin{align}
\notag  R_1&<R_{\text{SNNC}1}=\left\{
\begin{array}{ll}
\text{min}\left\{d_{31}, g_{31} \right\} & \text{if}\;D_{\text{SNNC}}\\
d_{33}& \text{otherwise}
\end{array}
\right.\\
\notag  R_2&<R_{\text{SNNC}2}=\left\{
\begin{array}{ll}
\text{min}\left\{e_{31}, h_{31} \right\} &\text{if}\;D_{\text{SNNC}}\\
e_{33 }&\text{otherwise}
\end{array}
\right.\\
R_1+R_2&<R_{\text{SNNC}3}=\left\{
\begin{array}{ll}
\text{min}\left\{f_{31}, i_{31}\right\}  & \text{if}\;D_\text{SNNC}\\
f_{33} &\text{otherwise}.
\end{array}
\right.
\end{align}
The event $D_\text{SNNC}$ means that the destination should decode the relay signal to achieve better performance.

The SNNC-DF rate region is the union of all $(R_1,R_2)$ satisfying
\begin{align}
 \notag  R_1<R_{\text{SNNC-DF}1}=\left\{
\begin{array}{ll}
R_{\text{DF}1}  & \text{if}\quad D_\text{DF}\\
R_{\text{SNNC}1} & \text{otherwise}
\end{array}
\right.\\
\notag  R_2<R_{\text{SNNC-DF}2}=\left\{
\begin{array}{ll}
R_{\text{DF}2}  & \text{if}\quad D_\text{DF}\\
R_{\text{SNNC}2} & \text{otherwise}
\end{array}
\right.\\
R_1 + R_2<R_{\text{SNNC-DF}3}=\left\{
\begin{array}{ll}
R_{\text{DF}3}  & \text{if}\quad D_\text{DF}\\
R_{\text{SNNC}3} & \text{otherwise}.
\end{array}
\right.
\end{align}
If $D_\text{DF}$ occurs, then the relay should decode which will remove interference at the relay. Otherwise, the relay should
perform QF to avoid unnecessarily lowering the rates. 

Let $R_{\text{tar}3}=R_{\text{tar}1} +R_{\text{tar}2}$. The outage probabilities are:
\begin{align}
&P^\text{out}_\text{DF}=\text{Pr}[\{R_{\text{DF}1}<R_{\text{tar}1}\} \cup \{R_{\text{DF}_2}<R_{\text{tar}2}\}  \cup
\{R_{\text{DF}_3}<R_{\text{tar}3}\}]\notag\\
&P^\text{out}_\text{CF-S}=\text{Pr}[\{R_{\text{CF-S}1}<R_{\text{tar}1} \}\cup \{R_{\text{CF-S}2}<R_{\text{tar}2}\}\notag\\
&\quad\quad\quad\quad\;\cup\{R_{\text{CF-S}}<R_{\text{tar}3}\}]\notag\\
&P^\text{out}_\text{SNNC}=\text{Pr}[\{R_{\text{SNNC}1}<R_{\text{tar}1}\} \cup \{R_{\text{SNNC}2}<R_{\text{tar}2}\}\notag\\ 
&\quad\quad\quad\quad\; \cup \{R_{\text{SNNC}3}<R_{\text{tar}3}\}]\notag\\
&P^\text{out}_\text{SNNC-DF}=\text{Pr}[\{R_{\text{SNNC-DF}1}<R_{\text{tar}1}\} \cup
\{R_{\text{SNNC-DF}2}<R_{\text{tar}2}\} \notag\\
&\quad\quad\quad\quad\;\cup \{R_{\text{SNNC-DF}3}<R_{\text{tar}3}\}]
\end{align}

\subsection{Two-Way Relay Channels}

\subsubsection{No Fading} 
The DF rate region for the Gaussian TWRC is the union of all $(R_1,R_2)$ satisfying
\begin{align}
R_1&<R_{\text{DF}1}=\text{min}\;\left\{a_{41}, a_{42} \right\}\notag\\
R_2&<R_{\text{DF}2}=\text{min}\;\left\{b_{41}, b_{42} \right\}\notag\\
R_1+R_2&<R_{\text{DF}3}=c_{41}\label{RDFTWRC}
\end{align}
where
\begin{align}
a_{41}&=C\left( |G_{13}|^2P_1(1-|\beta|^2)\right)\notag\\
a_{42}&=C\left(|G_{12}|^2P_1+|G_{32}|^2 P_3(1-|\theta_1|^2) \right.\notag\\
&\left.\quad\quad+2\Re\{\beta G_{12}( \theta_1 G_{32})^\ast\}\sqrt{P_1P_3}\right)\notag\\
b_{41}&=C\left(|G_{23}|^2P_2(1-|\gamma|^2)\right)\notag\\
b_{42}&=C\left(|G_{21}|^2P_2+|G_{31}|^2 P_3(1-|\theta_1|^2)\right.\notag\\ 
&\left.\quad\quad+2\Re\{\gamma G_{21}( \theta_2 G_{31})^\ast\}\sqrt{P_2P_3} \right)\notag\\
c_{41}&=C\left(|G_{13}|^2P_1(1-|\beta|^2)+|G_{23}|^2P_2(1-|\gamma|^2)\right)
\end{align}
where $0\le |\beta|^2, |\gamma|^2 \le1$ and $\sum^2_{i=1}|\theta_i|^2=1 $. The optimal power allocation
parameters are calculated numerically.

The CF-S rate region \cite[Proposition 4]{Rankov01} is the union of all $(R_1,R_2)$ satisfying 
\begin{align}
R_1&<d_{41}\notag\\
R_2&<e_{41}
\end{align}
where
\begin{align*}
d_{41}&=C\left(|G_{12}|^2P_1+\frac{|G_{13}|^2P_1}{1+\hat \sigma^2_3}\right)\\
e_{41}&=C\left(|G_{21}|^2P_2+\frac{|G_{23}|^2P_2}{ 1+\hat \sigma^2_3}\right)
\end{align*}
for some
\begin{align*}
\hat \sigma^2_3\ge \text{max}\;\left\{f_{41}, f_{42}, f_{43}, f_{44}\right\}
\end{align*}
where
\begin{align*}
f_{41}&=\frac{1+|G_{12}|^2P_1+ |G_{13}|^2P_1}{|G_{32}|^2P_3}\\
f_{42}&=\frac{|G_{21}|^2P_2+ 1}{|G_{31}|^2P_3}+\frac{|G_{13}|^2 P_1(|G_{21}|^2P_2+1)}{|G_{31}|^2P_3(|G_{12}|^2P_1+ 1)}\\
f_{43}&=\frac{1+|G_{21}|^2P_2+ |G_{23}|^2P_2}{|G_{31}|^2P_3},\\
f_{44}&=\frac{|G_{12}|^2P_1+ 1}{|G_{32}|^2P_3}+\frac{|G_{23}|^2 P_2 (|G_{12}|^2P_1+
1)}{|G_{32}|^2P_3(|G_{21}|^2P_2+ 1)}.
\end{align*}

Referring to Theorem~\ref{thm:theorem1}, the SNNC rate region is the union of all $(R_1,R_2)$
satisfying
\begin{align}
R_1&<\text{min}\left\{ d_{41}, g_{41}\right\}\notag\\
R_2&<\text{min}\left\{ e_{41}, h_{41}\right\}
\end{align}
where
\begin{align*}
g_{41}&=C\left(|G_{12}|^2P_1+|G_{32}|^2P_3\right)-C\left(\frac{1}{\hat \sigma^2_3}\right)\\
h_{41}&=C\left(|G_{21}|^2P_2 + |G_{31}|^2P_3\right)-C\left(\frac{1}{\hat \sigma^2_3}\right)
\end{align*}
for some $\hat \sigma^2_3 > 0$. The SNNC-DF rate region is the union of the DF and SNNC rate regions.


\subsubsection{Slow Rayleigh Fading}
Define the events
\begin{align*}
D_{\text{DF}}&=\left\{
\begin{array}{ll}
R_{\text{tar}_1}<a_{41}\\
R_{\text{tar}_2}<b_{41}\\
R_{\text{tar}_1}+R_{\text{tar}_2}<c_{41}
\end{array}
\right\}\\
D_{\text{CF-S}{11}}&=\left\{\;R_{3(\text{bin})}<C\left(\frac{|G_{31}|^2P_3}{1+ |G_{21}|^2P_2}\right)\right\}\\
D_{\text{CF-S}{12}}&=\left\{ R_{3(\text{bin})}  \ge C \left( \frac{1}{\hat \sigma^2_3} +
\frac{|G_{23}|^2P_2}{\hat \sigma^2_3 \left( 1+|G_{21}|^2P_2 \right) }\right)\right\}\\
D_{\text{CF-S}{21}}&=\left\{R_{3(\text{bin})}<C\left(\frac{|G_{32}|^2P_3}{1+|G_{12}|^2P_1}\right)\right\}\\
D_{\text{CF-S}{22}}&=\left\{R_{3(\text{bin})}\ge C\left( \frac{1}{\hat \sigma^2_3} +
\frac{|G_{13}|^2P_1}{\hat \sigma^2_3 \left( 1+|G_{12}|^2P_1 \right) }\right)\right\}\\
D_{\text{SNNC}1}&=\left\{\hat \sigma^2_3\ge \frac{1}{ |G_{32}|^2P_3}\right\}\\
D_{\text{SNNC}2}&=\left\{\hat \sigma^2_3\ge \frac{1}{|G_{31}|^2P_3}\right\}.
\end{align*}
The DF region is the union of all $(R_1,R_2)$ satisfying (\ref{RDFTWRC}). The CF-S region is the union of all $(R_1,R_2)$
satisfying
\begin{align}
&R_1<R_{\text{CF-S}1}=\left\{
\begin{array}{ll}
d_{41} & \text{if}\;D_{\text{CF-S}{21}}\cap D_{\text{CF-S}{22}}\\ 
d_{42} & \text{if}\;D_{\text{CF-S}{21}}\cap D^\text{c}_{\text{CF-S}{22}}\\
d_{43} & \text{otherwise}
\end{array}
\right.\\
&R_2<R_{\text{CF-S}2}=\left\{
\begin{array}{ll}
e_{41}  & \text{if}\: D_{\text{CF-S}{11}}\cap D_{\text{CF-S}{12}}\\ 
e_{42}  & \text{if}\: D_{\text{CF-S}{11}}\cap D^\text{c}_{\text{CF-S}{12}}\\
e_{43}  & \text{otherwise}
\end{array}
\right.
\end{align}
where
\begin{align*}
d_{42}&=C\left(|G_{12}|^2P_1 \right)\\
d_{43}&=C\left( \frac{|G_{12}|^2P_1 }{1+|G_{32}|^2P_3} \right)\\
e_{42}&=C\left(|G_{21}|^2P_2 \right) \\
e_{43}&=C\left( \frac{|G_{21}|^2P_2}{1+ |G_{31}|^2P_3} \right).
\end{align*}
The optimal $R_{3(\text{bin})}$ and $\hat \sigma^2_3$ are calculated numerically.

Referring to Theorem~\ref{thm:theorem1}, SNNC achieves all pairs $(R_1,R_2)$ satisfying
\begin{align}
&R_1<R_{\text{SNNC}1}=\left\{
\begin{array}{ll}
\text{min}\:\left\{d_{41}, g_{41}\right\} & \text{if}\; D_{\text{SNNC}1}\\
d_{43}  & \text{otherwise}
\end{array}
\right.\\
&R_2<R_{\text{SNNC}2}=\left\{
\begin{array}{ll}
\text{min}\:\left\{e_{41}, h_{41}\right\} & \text{if}\; D_{\text{SNNC}2}\\
e_{43}  & \text{otherwise}.
\end{array}
\right.
\end{align}
The SNNC-DF rate region is the union of the $(R_1, R_2)$ satisfying
\begin{align}
&R_1<R_{\text{SNNC-DF}1}= \left\{
\begin{array}{ll}
R_{\text{DF}1}  & \text{if}\: D_{\text{DF}}\\
R_{\text{SNNC}1} & \text{otherwise}
\end{array}
\right.\\
&R_2<R_{\text{SNNC-DF}2}=\left\{
\begin{array}{ll}
R_{\text{DF}2}  & \text{if}\: D_{\text{DF}}\\
R_{\text{SNNC}2} & \text{otherwise}.
\end{array}
\right.
\end{align}

The outage probabilities are:
\begin{align}
&P^\text{out}_\text{DF}=\text{Pr}[\{R_{\text{DF}1}<R_{\text{tar}1}\} \cup \{R_{\text{DF}_2}<R_{\text{tar}2}\}]\notag\\
&P^\text{out}_\text{CF-S}=\text{Pr}[\{R_{\text{CF-S}1}<R_{\text{tar}1}\} \cup\{ R_{\text{CF-S}2}<R_{\text{tar}2}\} ]\notag\\
&P^\text{out}_\text{SNNC}=\text{Pr}[\{R_{\text{SNNC}1}<R_{\text{tar}1}\} \cup \{ R_{\text{SNNC}2}<R_{\text{tar}2}\} ]\notag\\
&P^\text{out}_\text{SNNC-DF}=\text{Pr}[\{R_{\text{SNNC-DF}1}<R_{\text{tar}1} \}\cup \{ R_{\text{SNNC-DF}2}<R_{\text{tar}2}\} ]
\end{align}

\end{appendices}

\bibliographystyle{IEEEtran}

\bibliography{paper}

\end{document}